\documentclass[aps,prd,preprint,groupedaddress,preprintnumbers,longbibliography,nofootinbib]{revtex4-1}
\newcommand{\lb}{\left\lbrace}
\newcommand{\rb}{\right\rbrace}
\newcommand{\Op}{\mathcal{O}}
\newcommand{\op}[1]{\left[ #1 \right]}
\newcommand{\N}{\mathcal{N}}

\newcommand{\QQ}{\mathcal{Q}}

\newcommand{\vev}[1]{\left\langle #1 \right\rangle}
\newcommand{\vvev}[1]{\left\langle\kern-0.3em\left\langle #1
    \right\rangle\kern-0.3em\right\rangle}
\newcommand{\ep}{\epsilon}
\newcommand{\vep}{\varepsilon}

\usepackage{amsmath}
\usepackage{graphicx}
\usepackage{amsfonts}

\begin{document}

% Use the \preprint command to place your local institutional report
% number in the upper righthand corner of the title page in preprint mode.
% Multiple \preprint commands are allowed.
% Use the 'preprintnumbers' class option to override journal defaults
% to display numbers if necessary
%\preprint{Mainz preprint number}
\preprint{KOBE-TH-20-01}

%Title of paper
\title{Operator product expansion coefficients in the exact renormalization group formalism}

% repeat the \author .. \affiliation  etc. as needed
% \email, \thanks, \homepage, \altaffiliation all apply to the current
% author. Explanatory text should go in the []'s, actual e-mail
% address or url should go in the {}'s for \email and \homepage.
% Please use the appropriate macro foreach each type of information

% \affiliation command applies to all authors since the last
% \affiliation command. The \affiliation command should follow the
% other information
% \affiliation can be followed by \email, \homepage, \thanks as well.
\author{C.~Pagani}
\email[]{carlo.pagani@lpmmc.cnrs.fr}
%\homepage[]{Your web page}
%\thanks{}
%\altaffiliation{}
\affiliation{
Universit\'e Grenoble Alpes, CNRS, LPMMC, 25 avenue des Martyrs,
38000 Grenoble, France}
\affiliation{Institute f\"{u}r Physik (WA THEP)
  Johannes-Gutenberg-Universit\"{a}t\\ Staudingerweg 7, 55099 Mainz,
  Germany}
\author{H.~Sonoda}\email[]{hsonoda@kobe-u.ac.jp}\affiliation{Physics
  Department, Kobe University, Kobe 657-8501, Japan}

%Collaboration name if desired (requires use of superscriptaddress
%option in \documentclass). \noaffiliation is required (may also be
%used with the \author command).
%\collaboration can be followed by \email, \homepage, \thanks as well.
%\collaboration{}
%\noaffiliation

\date{\today}

\begin{abstract}
  We study how to compute the operator product expansion coefficients
  in the exact renormalization group formalism.  After discussing
  possible strategies,
  we consider some examples explicitly, within the $\ep$-expansions,
  for the Wilson-Fisher fixed points of the real scalar theory in
  $d=4-\ep$ dimensions and the Lee-Yang model in $d=6-\ep$ dimensions.
  Finally we discuss how our formalism may be extended beyond
  perturbation theory.
\end{abstract}

% insert suggested PACS numbers in braces on next line
% insert suggested keywords - APS authors don't need to do this
%\keywords{}

%\maketitle must follow title, authors, abstract, \pacs, and \keywords
\maketitle

\section{Introduction}

The exact renormalization group (ERG) provides a framework to study
the fundamental aspects of quantum field theories (QFT). For instance,
it allows one to give a non-perturbative definition of renormalizable
theories \cite{Wilson:1973jj} and to discuss the realization of
symmetries at the quantum level \cite{Igarashi:2009tj}. On top of
being a conceptual framework, ERG offers a framework for practical
computations.  It has been used as a computational tool for the
universal quantities such as critical exponents in statistical field
theory (see e.g.~\cite{Berges:2000ew,Delamotte:2007pf}), and also as
an exploratory tool in a wide range of subjects including quantum
gravity \cite{R96PaReut2018}.

Besides ERG, the operator product expansion (OPE) offers important
insights into non-perturbative aspects of QFTs \cite{Wilson:1969zs}. 
Let us denote
composite operators as $\op{\Op_{a}}$ and the product of two composite
operators as $\op{\Op_{a}\Op_{b}}$.  OPE states the validity of
\begin{align}
  \op{\Op_{a}\left(x\right)\Op_{b}\left(y\right)}
  & = \sum_{c}C_{abc}\left(x-y\right)\left[\Op_{c}\left(\frac{x+y}{2}\right)\right]
    \label{eq:OPE-statement}  
\end{align}
inserted into any correlation functions for small $|x-y|$. The
existence of OPE has been proved perturbatively by Zimmermann
\cite{Zimmermann:1972tv} and plays a fundamental role in the study of
conformal field theories (CFTs). Since both ERG and OPE offer
non-perturbative insights into QFT, it is natural to study the
relation between the two.

ERG was used to provide a perturbative proof of the existence of OPE
\cite{Hughes:1988cp,Keller:1991bz,Keller:1992by,Hollands:2011gf,Holland:2014ifa,Holland:2014pna},
but little effort had been made to explore OPE within ERG beyond
perturbation theory. In \cite{Pagani:2017tdr} a non-perturbative
definition of operator products was given, and simple examples of OPE
were constructed in the Wilson action framework.  Recently, composite
operators have been constructed explicitly in the ERG formalism to make
contact with various physical observables
\cite{Becchi:1996an,Igarashi:2009tj,Rose:2015bma,Rose:2016elj,Daviet:2018lfy,Pagani:2015hna,Pagani:2016dof,Becker:2018quq,Becker2019Frank2020}.

In the present work we construct explicit examples of OPEs within the
effective average action (EAA) framework \cite{Wetterich:1992yh}. We
identify two possible strategies to compute OPEs via ERG. One is based
on the construction of operator products and their expansions in
composite operators, and the other is based on the computation of
three-point functions for theories with conformal invariance. We study
explicit solutions of the ERG equations perturbatively with the
$\ep$-expansions, and compare our results with those obtained via the
conformal bootstrap
\cite{Gaiotto:2013nva,Hasegawa:2016piv,Gopakumar:2016wkt,Gopakumar:2016cpb,Gopakumar:2018xqi}.
We also comment on how we may extend our strategies to
non-perturbative approximation schemes available within the ERG
formalism.

The paper is organized as follows.  In section
\ref{sec:Operator-product-expansion-in-the-ERG} we define composite
operators and their products in the ERG formalism.  We then consider
the expansion of an operator product in composite operators and
outline two possible strategies for the calculation of such an
expansion.  In section \ref{sec:Wilson-Fisher-fp} we give technical
remarks to explain how to construct operators at the Wilson-Fisher
fixed point.  Discussions of a fixed point require fixing a momentum
cutoff, and we augment the technical outline given in section
\ref{sec:Operator-product-expansion-in-the-ERG} where the momentum
cutoff flows.  Section \ref{sec:Scaling-operators-from-ERG} is a long
technical section, where we construct some composite operators
explicitly and normalize them appropriately.  In section
\ref{sec:OPE-coefficients-from-ERG} we apply the results obtained
above to extract OPE coefficients and discuss our findings.  We
summarize our results in section \ref{sec:summary-and-outlook} and
discuss the future perspectives.

\section{Operator product expansion in the ERG formalism \label{sec:Operator-product-expansion-in-the-ERG}}

\subsection{Operator products in the ERG \label{sub:Operator-products-in-the-ERG}}

We extend the usual ERG formalism by introducing external sources that
couple to composite operators so that the correlation functions of
composite operators can be retrieved directly from the Wilson action.
In this paper we actually prefer to discuss the 1PI counterpart of the
Wilson action, known as the effective average action (EAA), so that we
only need to deal with the 1PI part of the correlation functions which
are relevant to the short distance singularities.  A fully analogous
construction applies to the Wilson action; see \cite{Pagani:2017tdr}
for example.
For an overview regarding composite operators in the ERG formalism, we
refer the reader to
\cite{Becchi:1996an,Pawlowski:2005xe,Igarashi:2009tj,Pagani:2016pad,Pagani:2017tdr}.

Let us consider the following modified generating functional of connected
correlation functions:
\begin{align*}
  Z_{k}\left[J,\vep\right]
  &= e^{W_{k}\left[J,\vep\right]} \\
  &\equiv  {\cal Z}\int{\cal D}\chi\,\exp\left[-S\left[\chi\right]+\int
  d^{d}x\,J\left(x\right)\chi\left(x\right)-\int
  d^{d}x\,\vep_{i}\left(x\right)\Op_{i}\left(x\right)-\Delta
  S_{k}\left[\chi\right]\right]\,, 
\end{align*}
where $S [\chi]$ is the bare action, and we have introduced the
sources $J (x), \vep_i (x)$ to the elementary field $\chi (x)$ and the
composite operator $\Op_{i}\left(x\right)$.  The momentum $k$ is an IR
cutoff, introduced via the IR cutoff action
$\Delta S_{k}\left[\chi\right]$ defined by
\[
  \Delta S_{k}\left[\chi\right] \equiv \int
  d^{d}x\,\frac{1}{2}\chi\,{\cal
    R}_{k}\left(-\partial^{2}\right)\chi\,.
\]
The kernel
$\mathcal{R}_{k}\left(-\partial^{2}\right)\equiv k^{2}R
\left(-\partial^{2}/k^{2}\right)$
suppresses the integration over the low momentum modes of $\chi$.

The EAA is then defined by the Legendre transform of
$W_{k}\left[J,\vep\right]$: 
\[
  \Gamma_{k}\left[\varphi,\vep\right]+\Delta
  S_{k}\left[\varphi\right] = \int
  d^{d}x\,J\left(x\right)\varphi\left(x\right)-W_{k}\left[J,\vep\right]\,,
\]
where
\[
  \varphi (x) =\frac{\delta W_{k} [J, \vep]}{\delta J (x)}\,,
\]
and no Legendre transform is performed over $\vep$. A composite
operator $\op{\Op_{a}}\left(x\right)$ is defined by
\begin{equation}
  \op{\Op_{a}}\left(x\right) \equiv
  \frac{\delta\Gamma_{k}\left[\varphi,\vep\right]}
  {\delta\vep_{a}\left(x\right)}\Bigr|_{\vep=0}\,.
  \label{eq:def-composite-operator}
\end{equation}
This is motivated by its correspondence with
$\frac{\delta}{\delta\vep}Z_{k}$ at fixed source $J$.  As a functional
of $\varphi$, $\op{\Op_a }(x)$ gives the 1PI part of the correlation
functions
\[
  \vev{\op{\Op_a} (x) \chi (x_1) \cdots \chi (x_n)}_S
\]
in the limit $k \to 0$.

\begin{figure}
\includegraphics[scale=0.2]{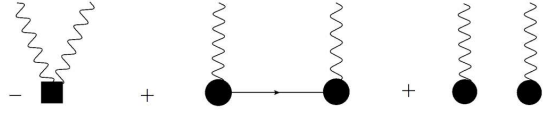}
\caption{Diagrammatic representation of the full correlation function
  $\left[\Op_{a}\Op_{b}\right]$ in the EAA formalism. The black dots
  denote composite operators $\op{\Op_a}, \op{\Op_b}$, i.e., the terms
  linear in the sources $\vep$.  The black square is the quadratic term
  $\frac{\delta^{2}\Gamma}{\delta\vep_a \delta\vep_b}\Bigr|_{\vep=0}$.
  \label{fig:Diagrammatic-representation-of-full-OO}}
\end{figure}

The product $\left[\Op_{a}\Op_{b}\right]\left(x,y\right)$ of two operators
$\op{\Op_a} (x), \op{\Op_b} (y)$ is defined by
\begin{align}
  \left[\Op_{a}\Op_{b}\right]\left(x,y\right)
  &\equiv -\frac{\delta^{2}\Gamma_k \left[\varphi,\vep\right]}
    {\delta\vep_{a}\left(x\right)\delta\vep_{b}\left(y\right)}\Bigr|_{\vep=0}  
    +\int_{z_{1},z_{2}} \frac{\delta^{2} \Gamma_{k} [\varphi,
    \vep]}{\delta \vep_{b} 
    \left(x\right)\delta\varphi \left(z_{1}\right)}\Bigr|_{\vep=0} \cdot 
    G_{k}\left(z_{1},z_{2}\right) \cdot \frac{\delta^{2}
    \Gamma_{k}  [\varphi, \vep]}{\delta \varphi \left(z_{2}\right)
    \delta\vep_{b} \left(y\right)}\Bigr|_{\vep=0}\nonumber 
  \\ 
  & \quad +\frac{\delta \Gamma_{k} [\varphi, \vep]}{\delta \vep_{a}
    \left(x\right)}\Bigr|_{\vep=0} \frac{\delta \Gamma_{k} [\varphi,
    \vep]}{\delta 
    \vep_{b} \left(y\right)} \Bigr|_{\vep=0}\,,\label{eq:def-of-operator-prod} 
\end{align}
where $G_{k}\left(z_{1},z_{2}\right) [\varphi]$ is the field dependent
high-momentum propagator defined by
\begin{equation}
\int d^d x\,
  \frac{\delta^2 \Gamma_k [\varphi, 0]}{\delta \varphi (z_1) 
    \delta \varphi (x)} G_k \left(x, z_2\right)
+  \mathcal{R}_k \left(- \partial_{z_1}^2\right) G_k \left(z_1,z_2\right)
 = \delta (z_1-z_2)\,.
  \end{equation}
  The product $\left[\Op_{a}\Op_{b}\right]\left(x,y\right)$ can be
  represented diagrammatically as in Fig.
  \ref{fig:Diagrammatic-representation-of-full-OO} which is motivated
  by its correspondence with
  $\frac{\delta^{2}}{\delta\vep_a \delta\vep_b}Z_{k}$ at fixed source
  $J$.\footnote{The connected part of the product is given by
\[
  \frac{\delta^{2}}{\delta \vep_{b} \left(x_b\right) \delta \vep_{a}
  \left(x_a\right)}\Bigr|_{J}W
  =  -\frac{\delta^{2} \Gamma \left[\varphi,\vep\right]}{\delta \vep_{a}
    \left(x_a\right) \delta\vep_{b} \left(x_b\right)}
    +\int_{x,y}\frac{\delta^{2} \Gamma [\varphi,
      \vep]}{\delta\vep_{b}\left(x_b\right)  
    \delta\varphi \left(x\right)} G_k \left(x,y\right) \frac{\delta^{2}
    \Gamma \left[\varphi,\vep\right]}{\delta\varphi \left(y\right)
    \delta\vep_{a}\left(x_a\right)}\,, 
\]
where we have used
\[
\frac{\delta}{\delta\vep_{c}\left(x_c\right)}\Bigr|_{J}  =
\frac{\delta}{\delta\vep_{c} \left(x_c\right)}\Bigr|_{\varphi}
-\int_{x,y} \frac{\delta^{2} \Gamma [\varphi, \vep]}{\delta \vep_{c}
  \left(x_c\right) 
  \delta\varphi \left(x\right)} G \left(x,y\right)
\frac{\delta}{\delta\varphi \left(y\right)}\,.
\]
}
See also \cite{Rose:2015bma}.

Both $\left[\Op_{a}\right]$ and $\left[\Op_{a}\Op_{b}\right]$ can be
constructed from the associated ERG, as summarized below. The EAA
$\Gamma_{k}\left[\varphi,\vep\right]$ satisfies the ERG differential
equation
\cite{Wetterich:1992yh,Ellwanger:1993mw,Morris:1993qb,Reuter:1993kw}
\begin{equation}
  \partial_{t}\Gamma_{k}\left[\varphi,\vep\right]
   =\frac{1}{2}\mbox{Tr}\left[\left(\Gamma_{k}^{\left(2\right)}
    \left[\varphi,\vep\right]+{\cal R}_{k}\right)^{-1}
    \partial_{t}{\cal R}_{k}\right]\,,\label{eq:ERG-EAA-full} 
\end{equation}
where
\begin{equation}
  \partial_t \equiv - k \frac{\partial}{\partial k}\,,\label{eq:log-flow-time}
\end{equation}
and
$\Gamma_{k}^{\left(2\right)}\left[\varphi,\vep\right] = \frac{\delta^2
  \Gamma_k}{\delta \varphi \delta \varphi}$
is the Hessian of the EAA with respect to the average field. By taking
the functional derivative of Eq.~(\ref{eq:ERG-EAA-full}) we can
derive the flow equations for composite operators and for the 1PI
contributions to the operator products
(\ref{eq:def-of-operator-prod}). More precisely, we obtain
\cite{Pawlowski:2005xe,Igarashi:2009tj,Pagani:2016pad}
\begin{equation}
  \partial_{t}\left[\Op_{a}\right]\left(x\right)
   =  -\frac{1}{2}\mbox{Tr} \left[\left(\Gamma_{k}^{\left(2\right)} +
    {\cal R}_{k}\right)^{-1} \left[\Op_{a}\right]^{\left(2\right)}
    \left(x\right)\left(\Gamma_{k}^{\left(2\right)} + {\cal
    R}_{k}\right)^{-1} \partial_{t}{\cal R}_{k}\right]\,,
    \label{eq:ERG-composite-operators}
\end{equation}
where
$\left[\Op_{a}\right]^{\left(2\right)} = \frac{\delta^2}{\delta \varphi
  \delta \varphi} \op{\Op_a}$ is the Hessian of the composite
operator. For the products, we obtain
\begin{align}
  & \partial_{t}\frac{\delta^{2}\Gamma_k \left[\varphi,\vep\right]}
  {\delta\vep_{a}\left(x\right) \delta\vep_{b} \left(y\right)}\Bigr|_{\vep=0}
  =  -\frac{1}{2}\mbox{Tr}\left[\left(\Gamma_{k}^{\left(2\right)} +
    {\cal R}_{k}\right)^{-1} \frac{\delta^{2} \Gamma_k^{\left(2\right)}
    \left[\varphi,0\right]}{\delta \vep_{a} \left(x\right)
    \delta\vep_{b} \left(y\right)}\left(\Gamma_{k}^{\left(2\right)} +
    {\cal R}_{k}\right)^{-1} \partial_{t}{\cal
    R}_{k}\right] \label{eq:ERG-1PI-operator-product} \\ 
 & \quad +\mbox{Tr}\left[\left( \Gamma_{k}^{\left(2\right)} + {\cal
   R}_{k}\right)^{-1} \frac{\delta \Gamma_k^{\left(2\right)}
   \left[\varphi,0\right]}{\delta
   \vep_{a}\left(x\right)}\left(\Gamma_{k}^{\left(2\right)} + {\cal
   R}_{k}\right)^{-1} \frac{\delta\Gamma_k^{\left(2\right)}
   \left[\varphi,0\right]}{\delta \vep_{b}
   \left(y\right)}\left(\Gamma_{k}^{\left(2\right)} + {\cal
   R}_{k}\right)^{-1}\partial_{t}{\cal R}_{k}\right]\,. \nonumber 
\end{align}
By solving Eqs. (\ref{eq:ERG-composite-operators}) and
(\ref{eq:ERG-1PI-operator-product}), we can construct the operator
product $\left[\Op_{a}\Op_{b}\right]$ as in
(\ref{eq:def-of-operator-prod}).

\subsection{Operator product expansion coefficients in the ERG
  formalism \label{sub:Operator-product-expansion-coeff-in-the-ERG}}

The operator product expansion (\ref{eq:OPE-statement}) and its
connection to ERG has already been studied in the literature
\cite{Hughes:1988cp,Keller:1991bz,Keller:1992by,Hollands:2011gf,Holland:2014ifa,Holland:2014pna,Pagani:2017tdr}.
The ERG framework provides a further perturbative proof of the
existence of the OPE
\cite{Hughes:1988cp,Keller:1991bz,Keller:1992by,Hollands:2011gf,Holland:2014ifa,Holland:2014pna}.
Actually, in the ERG formalism, the OPE (\ref{eq:OPE-statement})
amounts to expressing the operator product
$\left[\Op_{a}\Op_{b}\right]\left(x,y\right)$, given by
(\ref{eq:def-of-operator-prod}), as an expansion in a basis of
composite operators, given by (\ref{eq:def-composite-operator}).
Thanks to the built-in locality, it has been argued that such an
expansion is natural in the ERG formalism \cite{Pagani:2017tdr}.

Let us emphasize that Eq. (\ref{eq:OPE-statement}) is expected to hold
along the entire RG trajectory.  When the theory is at criticality,
however, we expect
\begin{equation}
  \Bigr\langle\left[\Op_{a}\right]\left(x\right) \left[\Op_{a}\right]
  \left(0\right)\Bigr\rangle =
  \frac{1}{\left|x-y\right|^{2\Delta_{a}}}
\end{equation}
to be valid at \textbf{large} distances if the operators are
normalized properly.  Then, the OPE coefficients in
(\ref{eq:OPE-statement}) are given by
\begin{equation}
C_{abc}\left(x-y\right)  =
\frac{c_{abc}}{\left|x-y\right|^{\Delta_{a}+\Delta_{b}-\Delta_{c}}}\,,
\end{equation}
where $\Delta_{a}$ is the scale dimension of the operator $\op{\Op_a}$, and
$c_{abc}$ is a numerical factor.  We refer to this numerical factor
$c_{abc}$ as an \textbf{OPE coefficient}.  Theories with conformal symmetry are
the \textbf{IR limit} of such critical theories, and the OPE coefficient
$c_{abc}$ appears as the overall coefficient of the three-point
function
\begin{equation}
  \Bigr\langle \op{\Op_{a}}\left(x_{1}\right) \op{\Op_{b}}\left(x_{2}\right)
  \op{\Op_{c}}\left(x_{3}\right)\Bigr\rangle  = 
  \frac{c_{abc}}{\left(x_{12}^{2}\right)^{d/2-\nu_{c}}
    \left(x_{23}^{2}\right)^{d/2-\nu_{a}}
    \left(x_{12}^{2}\right)^{d/2-\nu_{b}}}\,,
  \label{eq:CFT-three-point-function} 
\end{equation}
where $x_{ij}^{2}=\left|x_{i}-x_{j}\right|^{2}$, $c_{abc}$ is fully
symmetric in its indices, and
$\nu_{a}=\frac{1}{2}\left(d+\Delta_{a}-\Delta_{b}-\Delta_{c}\right)$,
etc.

The main aim of this paper is to lay down a possible strategy for
computing the OPE coefficients $c_{ijk}$ within the ERG framework. We
can consider the following two possibilities.
\begin{description}
\item [{A)}] Construct the full operator product
  $\left[\Op_{a}\Op_{b}\right]$ first, and expand it in a basis of
  composite operators $\left[\Op_{c}\right]$.
\item [{B)}] Assuming conformal symmetry, extract the coefficient
  $c_{abc}$ directly from the three-point function given by
  (\ref{eq:CFT-three-point-function}).
\end{description}
The strategy {\bf A)} has been followed in \cite{Pagani:2017tdr} to
provide simple examples around the Gaussian fixed point. Although
available in principle, this strategy is cumbersome in practice. Since
one usually works in momentum space, it is easier to construct the
most singular (i.e., non-integrable) part of the OPE rather than the
full (i.e., including the finite part) OPE.

In this paper we will consider the approach {\bf B)} in detail within
the ERG formalism.
%Even though our example can be derived also in the
%{\bf A)}-approach, as we check explicitly, we are going to compute
We will extract OPE coefficients from the associated three-point
functions.  We calculate in momentum space rather than in coordinate
space.  In subsection \ref{sub:OPE-coeff-in-momentum-space} below we
recall a few basic features of conformal invariance in momentum space
and explain the main points of our recipe.  As a side note, we point
out that ERG is a very efficient framework to discuss the presence of
conformal invariance
\cite{Delamotte:2015aaa,Delamotte:2018fnz,DePolsi:2018vxc,DePolsi:2019owi,Rosten:2016nmc,Rosten:2017urs,Sonoda:2017zgl}.

We apply the strategy outlined above to compute some examples of the
OPE coefficients. We solve the ERG differential equations
perturbatively and express the results via the $\ep$-expansion. This
allows us to compare our results with others that have been obtained
in the literature by means of the analytic conformal bootstrap
approach
\cite{Gaiotto:2013nva,Hasegawa:2016piv,Gopakumar:2016wkt,Gopakumar:2016cpb,Gopakumar:2018xqi}.
In particular, we will compute the coefficients ``$c_{112}$'' and
``$c_{114}$'' to the first order in $\ep$ for the critical Ising model
in $d=4-\ep$ dimensions, and the coefficient ``$c_{111}$'' to the
first order in $\sqrt{\ep}$ for the critical Lee-Yang model in
$d = 6-\ep$ dimensions. Our results agree with those obtained by other
approaches.

\subsection{Operator product expansion coefficients in momentum space \label{sub:OPE-coeff-in-momentum-space}}

Calculations in quantum field theory often rely on momentum space,
where the loop corrections are most easily expressed and computed.  On
the other hand, formulas regarding OPE are often most naturally
expressed in real (coordinate) space, especially in the case of
CFTs. In this subsection we briefly describe the connection between
expressions in real space and those in momentum space.

Conformal symmetry in momentum space has been analyzed in the
literature in some detail; see
\cite{Coriano:2013jba,Maglio:2019grh,Bzowski:2013sza,Bzowski:2015pba,Bzowski:2015yxv,Bzowski:2019kwd,Isono:2019ihz,Gillioz:2018mto}
and references therein.  We only need a few very basic formulas, which
we review below (a few more technical details are provided in Appendix
\ref{App:Momentum-space-2pt-3pt}).  We refer the interested reader to
the literature given above for more details regarding the constraint
imposed by conformal symmetry in momentum space.

Let us consider a set of primary operators $\phi_{i}$, normalized by
the two-point functions
\begin{equation}
\vev{\phi_{i} \left(x\right) \phi_{j}\left(0\right)} =
\frac{1}{x^{\Delta_{i}+\Delta_{j}}} \delta_{ij}\,.
\label{eq:CFT-2pt-function-real-space}
\end{equation}
In momentum space this normalization gives
\begin{equation}
  \vev{\phi_{i}\left(p\right)\phi_{i}\left(q \right)} = (2\pi)^d
  \delta (p+q) \cdot \pi^{d/2}
  \frac{\Gamma \left(\frac{d}{2}-\Delta_{i}\right)}{\Gamma
    (\Delta_{i})}
  \left(\frac{4}{p^{2}}\right)^{\frac{d}{2}-\Delta_{i}}\,,
  \label{eq:CFT_2pt-function_momentum_space}
\end{equation}
where we have used (\ref{eq:Four-of-normalized-2pt-function}) given in
Appendix \ref{App:Momentum-space-2pt-3pt}.

The Fourier transform of the three-point function
\begin{equation}
  \vev{\phi_1 (x_1) \phi_2 (x_2) \phi_3 (x_3)} =
  \frac{c_{123}}{(x_{12}^2)^{d/2-\nu_3} (x^2_{23})^{d/2-\nu_1}
    (x^2_{12})^{d/2-\nu_2}}
\end{equation}
(the same as (\ref{eq:CFT-three-point-function})) is given by a
somewhat complicated expression either as a momentum integral or via
special functions. To avoid such complications, we consider the limit
$p_{1}\gg p_{2}$, where we find
\begin{align}
&\vev{\phi_{1}\left(p_{1}\right) \phi_{2}\left(p_{2}\right)
  \phi_{3}\left(-p\right)} \overset{p_1 \gg p_2}{\longrightarrow}
  (2\pi)^d \delta (p_1+p_2-p) \cdot 
  \frac{c_{123}}{p_{1}^{d+\Delta_{2}-\Delta_{1}-\Delta_{3}}
  p_{2}^{d-2\Delta_{2}}}\nonumber\\
  &\quad \times \left(4\pi\right)^{d}4^{-\frac{1}{2}\left(\Delta_{1} +\Delta_{2} +\Delta_{3}\right)}
  \frac{\Gamma\left(\frac{1}{2}
  \left(d+\Delta_{2}-\Delta_{1}-\Delta_{3}\right) \right)}{\Gamma
  \left(\frac{1}{2} \left(\Delta_{1}+\Delta_{3}-\Delta_{2}\right) \right)}
  \frac{\Gamma \left(\frac{d}{2}-\Delta_{2}\right)}{\Gamma
  \left(\Delta_{2}\right)}\,. 
  \label{eq:CFT_3pt-function_momentum_space} 
\end{align}
To extract the coefficient $c_{123}$, it is enough to know this
asymptotic behavior .  We provide a little more details of this
formula in Appendix \ref{App:Momentum-space-2pt-3pt}.

\section{Wilson-Fisher fixed point\label{sec:Wilson-Fisher-fp}}

In this section we wish to discuss technicalities that are quite
important for our calculations.  In the main part of this paper we
consider a real scalar theory at its criticality in $d = 4 - \ep$
dimensions.  In doing so with ERG, we have two alternatives:
\begin{description}
\item[{C)}] We start from a bare theory $S_B [\chi]$ with a large but
  finite cutoff $k_0$.  We tune the bare parameters to make the theory
  critical.  We then construct the EAA $\Gamma_k [\varphi, \vep]$
  whose $k$-dependence is determined by ERG.  In the limit $k \to 0$,
  $\Gamma_k$ becomes the 1PI generating functional of the correlation
  functions.  A conformal field theory is obtained as the \textbf{IR
    limit} of the critical theory, i.e., we must look at the
  correlation functions for the momenta much \textbf{smaller} than the
  bare cutoff $k_0$.
\item[{D)}] We adopt the dimensionless convention by measuring all
  dimensionful quantities in appropriate powers of the cutoff $k$.
  The resulting 1PI EAA $\bar{\Gamma}_t [\bar{\varphi}, \bar{\vep}]$
  has a fixed cutoff of order $1$, and satisfies an ERG differential
  equation with the Gaussian and Wilson-Fisher fixed-point solutions.
  $\bar{\Gamma}_* [\bar{\varphi}, \bar{\vep}]$ at the Wilson-Fisher
  fixed point gives the correlation functions of a conformal field
  theory, but only for the momenta much \textbf{larger} than the fixed
  cutoff.\cite{Sonoda:2017rro}
\end{description}

We prefer \textbf{D)} because it is easier to construct a fixed point
than to tune a bare theory for criticality.  For completeness and the
convenience of the reader, let us rewrite the relevant ERG
differential equations in the dimensionless convention.

We first introduce dimensionless fields by
\begin{subequations}
\begin{align}
  \bar{\varphi} (p) &\equiv k^{\frac{d+2}{2}} \varphi (p k)\,,\\
  \op{\bar{\Op}_i} (p) &\equiv k^{d-d_i} \op{\Op_i} (p k)\,,\\
  \bar{\vep}_i (p) &\equiv k^{d_i} \vep_i (p k)\,,
\end{align}
\end{subequations}
where $d_i$ is the scale dimension of
$\op{\Op_i} (x) = \int_p e^{i p x} \op{\Op_i} (p)$ in the coordinate
space at the Gaussian fixed point.  It is sometimes called the
engineering dimension.

We then define
\begin{equation}
  \bar{\Gamma}_t [\bar{\varphi}, \bar{\vep}] \equiv \Gamma_k [\varphi,
  \vep]\,,
\end{equation}
where we have traded the momentum cutoff $k$ for the logarithmic flow
time $t$, given by (\ref{eq:log-flow-time}).  The ERG differential
equation for $\bar{\Gamma}_t [\bar{\varphi}, \bar{\vep}]$ is given by
\begin{align}
  \partial_t \bar{\Gamma}_t [\bar{\varphi}, \bar{\vep}]
  &= \int_p \left(\frac{d+2}{2} + p \cdot \partial_p \right)
    \bar{\varphi} (p) \cdot \frac{\delta \bar{\Gamma}_t [\bar{\varphi},
    \bar{\vep}]}{\delta \bar{\varphi} (p)}
 + \int_p \left(d_i + p \cdot \partial_p \right) \bar{\vep}_i
    (p) \cdot \frac{\delta \bar{\Gamma}_t [\bar{\varphi},
    \bar{\vep}]}{\delta \bar{\vep}_i (p)}\notag\\
  &\quad - \int_p \left(2 - p \cdot \partial_p\right) R(p) \cdot
    \frac{1}{2} G_{t; p,-p} [\bar{\varphi}, \bar{\vep}]\,,
    \label{eq:ERG-Gammabar-full}
\end{align}
where $G_{t; p,-q} [\bar{\varphi}, \bar{\vep}]$ is defined by
\begin{equation}
  \int_q G_{t; p,-q} [\bar{\varphi}, \bar{\vep}] \left( \frac{\delta^2
      \bar{\Gamma}_t [\bar{\varphi}, \bar{\vep}]}{\delta \bar{\varphi}
      (q) \delta \bar{\varphi} (-r)} + R(q)\,(2 \pi)^d \delta (q-r) \right)
  = (2\pi)^d \delta (p-r)\,.
\end{equation}
The cutoff function $R(q)$ is a fixed function.

We are interested in the ERG flow from the Gaussian fixed point to the
Wilson-Fisher fixed point.  We parametrize the flow by $g$ so that the
Gaussian fixed point is at $g=0$, and the Wilson-Fisher fixed point is
at $g=g_*$.  We can then replace $\partial_t$ by
\[
  \left( \ep g + \beta (g) \right) \partial_g
\]
where $\ep$ is the scale dimension of $g$ at $g=0$.  Accordingly,
$\bar{\Gamma} (g) [\bar{\varphi}] = \bar{\Gamma}_t [\bar{\varphi}, 0]$
satisfies the ERG equation
\begin{equation}
  \lb \left( \ep g + \beta (g) \right) \partial_g +
  \Delta_{\bar{\varphi}} \rb \bar{\Gamma} (g)
  [\bar{\varphi}] = - \int_p \left(2 - p \cdot \partial_p\right) R(p) \cdot
  \frac{1}{2} G (g)_{p,-p} [\bar{\varphi}]\,,
  \label{eq:ERG-dimless-Gamma}
\end{equation}
where we define $\Delta_{\bar{\varphi}}$ and $G (g)_{p,-q}
[\bar{\varphi}]$ by
\begin{equation}
  \Delta_{\bar{\varphi}} \equiv \int_p \left( - \frac{d+2}{2} - p \cdot \partial_p
  \right) \bar{\varphi} (p) \cdot \frac{\delta}{\delta \bar{\varphi}
    (p)}\,,
\end{equation}
and
\begin{equation}
  \int_q G (g)_{p,-q} [\bar{\varphi}] \left( \frac{\delta^2 \bar{\Gamma} (g)
      [\bar{\varphi}]}{\delta \bar{\varphi} (q) \delta \bar{\varphi}
      (-r)} + R(q)\, (2 \pi)^d \delta (q-r) \right) = (2\pi)^d \delta
  (p-r)\,.
\end{equation}

Actually, for the Wilson-Fisher fixed point to exist at
$g=g_* = \mathrm{O} (\ep)$, we need to introduce an anomalous
dimension $\frac{1}{2} \eta (g) = \mathrm{O} (g^2)$ of
$\bar{\varphi}$.  Since we are only interested in the corrections of
order $\ep$ to the OPE coefficients, we can ignore it.  In Appendix
\ref{App:The-fixed-point-EAA-1st-order} we obtain $\beta (g)$ to order
$g^2$ to determine the fixed-point value
\begin{equation}
  g_* = \frac{(4 \pi)^2}{3} \ep + \mathrm{O} (\ep^2)\,.\label{eq:fixed-point-value}
\end{equation}

Similarly, differentiating (\ref{eq:ERG-Gammabar-full}) with respect
to $\bar{\ep}_i$, we obtain the ERG equation for $\op{\bar{\Op}_i}$ as
\begin{align}
  &\lb \left( \ep g + \beta (g) \right) \partial_g  + d - d_i + p
    \cdot \partial_p + \Delta_{\bar{\varphi}} \rb  \op{\bar{\Op}_i} (p)
    - \sum_j \gamma_{ij} (g) 
    \op{\bar{\Op}_j} (p) \notag\\
    &= \int_q (2 - q \cdot \partial_q) R(q)
    \,\frac{1}{2} \int_{r,s} G (g)_{q,-r} [\bar{\varphi}]
    \frac{\delta^2 \op{\bar{\Op}_i} 
    (p)}{\delta \bar{\varphi} (r) \delta \bar{\varphi} (s)}
    G(g)_{-s,-q} [\bar{\varphi}]\,.\label{eq:dimless-ERG-Oi}
\end{align}
The mixing matrix $\gamma_{ij} (g)$ results from appropriate boundary
conditions imposed at $p=0$.  Differentiating
(\ref{eq:ERG-Gammabar-full}) once more, we obtain the ERG equation for
\begin{equation}
  \QQ_{ij} (p_1, p_2) \equiv  - \frac{\delta^2 \bar{\Gamma} (g)
    [\bar{\varphi}, \bar{\vep}]}{\delta \bar{\vep}_i (-p_1) \delta
    \bar{\vep}_j (-p_2)}\Bigr|_{\vep=0}
\end{equation}
as
\begin{align}\label{eq:dimless-ERG-Qij}
  & \lb \left( \ep g + \beta(g) \right) \partial_g  + 2 d - d_i - d_j + p_1
    \cdot \partial_{p_1} + p_2 
    \cdot \partial_{p_2} + \Delta_{\bar{\varphi}} \rb\QQ_{ij}
    (p_1, p_2)\notag\\
  &\quad - \sum_k \left(\gamma_{ik}(g) \QQ_{kj} (p_1,p_2)+ \gamma_{jk} (g)
    \QQ_{ik} (p_1,p_2)+  \gamma_{ij,k} (g) \op{\bar{\Op}_k} (p_1+p_2)
    \right)\notag  \\ 
  &= \int_q \left(2 - q \cdot \partial_q\right) R(q)\,
    \int_{r,s} G(g)_{q,-r} [\bar{\varphi}] G(g)_{s,-q}
    [\bar{\varphi}]  \notag\\
  &\quad \times \left(
    \frac{1}{2} \frac{\delta^2 \QQ_{ij} (p_1, p_2)}{\delta \bar{\varphi} (r) \delta
    \bar{\varphi} (-s)} +\int_{u,v} \frac{\delta^2
    \op{\bar{\Op}_i} (p_1)}{\delta 
    \bar{\varphi} (r) \delta \bar{\varphi} (-u)} G(g)_{u,-v}
    [\bar{\varphi}] \frac{\delta^2 \op{\bar{\Op}_j} (p_2)}{\delta
    \bar{\varphi} (v) \delta \bar{\varphi} (-s)}\right)\,,
\end{align}
where the mixing $\gamma_{ij,k} (g)$ is due to appropriate boundary
conditions imposed at $p_1 = p_2 = 0$.\footnote{
Eqs.~(\ref{eq:dimless-ERG-Oi}) and (\ref{eq:dimless-ERG-Qij}) are derived by employing renormalized sources.}

We have thus given the defining ERG differential equations for
$\bar{\Gamma} (g) [\bar{\varphi}]$, $\op{\bar{\Op}_i} (g;p)$, and
$\QQ_{ij} (g;p_1,p_2)$.  We end this section by summarizing how to
extract the two- and three-point functions out of
these \cite{Sonoda:2017rro}. We first set $g$ to the fixed-point value
$g_*$ to go to the Wilson-Fisher fixed point.  To extract the
two-point function of the elementary scalar field, we expand
\begin{equation}
  \bar{\Gamma} (g_*) [\bar{\varphi}] = \frac{1}{2} \int_{p_1, p_2}
  \bar{\varphi} (p_1) \bar{\varphi} (p_2)\, (2\pi)^d \delta (p_1+p_2)
  \,\bar{\Gamma}^{(2)} (p_1) + \cdots 
\end{equation}
in powers of fields.  For $p \gg 1$, we obtain the two-point function as
\begin{equation}
  \vev{\bar{\varphi} (p) \bar{\varphi} (q)} = \frac{1}{\bar{\Gamma}^{(2)} (p)}\,
  (2\pi)^d \delta (p+q)\,.
\end{equation}
Similarly, the field independent part of
\begin{equation}
  \op{\bar{\Op}_i \bar{\Op}_j} (g_*; p_1, p_2) = \op{\bar{\Op}_i \bar{\Op}_j}^{(0)}
  (p_1)\, (2\pi)^d \delta (p_1+p_2) + \cdots
\end{equation}
gives, for $p_1, p_2 \gg 1$, the two-point function as
\begin{equation}
  \vev{\bar{\Op}_i (p_1) \bar{\Op}_j (p_2)} = \op{\bar{\Op}_i \bar{\Op}_j}^{(0)}
  (p_1)\, (2\pi)^d \delta (p_1+p_2) \,.
\end{equation}
Finally, the part quadratic in fields of
\begin{equation}
  \op{\bar{\Op}_i } (g_*; p) = \frac{1}{2} \int_{p_1, p_2} \bar{\varphi} (p_1)
  \bar{\varphi} (p_2) \, (2\pi)^d \delta (p_1+p_2-p)\,
  \op{\bar{\Op}_i}^{(2)} (p_1, p_2) + \cdots
\end{equation}
gives, for $p_1, p_2 \gg 1$, the three-point function as
\begin{equation}
  \vev{\bar{\varphi} (-p_1) \bar{\varphi} (-p_2) \bar{\Op}_i (p)}
  =   \op{\bar{\Op}_i}^{(2)} (p_1, p_2) \frac{1}{\bar{\Gamma}^{(2)} (p_1)
    \bar{\Gamma}^{(2)} (p_2)}\, (2\pi)^d \delta (p_1+p_2-p)
    \label{eq:formula-three-point}
\end{equation}
for $\mathbb{Z}_2$-invariant operators $\op{\bar{\Op}_i}$.  
This equation needs to be modified
as (\ref{eq:3-pt-function-general}) if the fixed point has no $\mathbb{Z}_2$ invariance.

In the remaining part of the paper we work only in the dimensionless
convention.  Hence, we omit the bars above the symbols altogether.
  
\section{Scaling operators from
  ERG \label{sec:Scaling-operators-from-ERG}}

In this section we construct explicitly scaling composite operators
$\phi_{i}$ at the Wilson-Fisher fixed point in $d=4-\ep$ dimensions by
solving the ERG. Composite operators are solutions of
(\ref{eq:dimless-ERG-Oi}), which we write again as
\begin{align}
&  \left( p \cdot \partial_p + \Delta_\varphi \right) \op{\Op_i} (p) +
  \left( d\, \delta_{ij} -  \Delta_{ij} \right) \op{\Op_j} (p)\notag\\
  &= \frac{1}{2} \int_q (2-q\cdot\partial_q) R(q) 
    \int_{r,s} G_{q,-r} [\Phi] G_{-q,-s} [\Phi] \frac{\delta^2 \op{\Op_i}
    (p)}{\delta \varphi (r) \delta \varphi (s)}\,,
    \label{eq:composite-op-at-FP}
\end{align}
where $G_{p,q} [\Phi]$ is defined by
\[
  \int_q G_{p, q} [\Phi] \left( \frac{\delta^2 \Gamma_*
      [\varphi]}{\delta \varphi (-q) \delta \varphi (r)} + R (q) (2
    \pi)^d \delta (q-r) \right) = (2 \pi)^d \delta (p-r)\,,
\]
and
\[
  \Delta_\varphi \equiv \int_q \left( - \frac{d+2}{2} - q
    \cdot \partial_q \right) \varphi (q) \cdot \frac{\delta}{\delta
    \varphi (q)}\,.
\]
$\Delta_{ij}$ is the matrix of scale dimensions, and it is not
diagonal in general:
\[
  \Delta_{ij}=d_{i}\,\delta_{ij}+\gamma_{ij}\,,
\]
where $d_i$ is the engineering (mass) dimension of $\Op_i$ in coordinate
space, and $\gamma_{ij}$ is the mixing matrix.

The scaling composite operators $\phi_{i}$ are suitable linear
combinations of composite operators that diagonalize the mixing matrix
$\Delta_{ij}$.  The scaling operators $\op{\Phi_i}$ for the eigenvalue $\Delta_i$ 
satisfy
\begin{align}
&\left(p \cdot \partial_p + d - \Delta_i + \Delta_{\varphi} \right)\left[\Phi_i \right] (p)\notag\\
&= \frac{1}{2} \int_q (2-q \cdot \partial_q) R(q) \int_{r,s}
  G_{q,-r} [\varphi] G_{-q,-s} [\varphi] \frac{\delta^2 \op{\Phi_i} (p)}{\delta
  \varphi (r) \delta \varphi (s)}\,.
\label{eq:scaling-composite-op-at-FP}
\end{align}
We can then introduce a normalization constant $\N_i$ so that
$\phi_{i} \equiv{\cal N}_i \left[\Phi_i \right]$ satisfies the
normalization (\ref{eq:CFT-2pt-function-real-space}). Note that the
coefficient ${\cal N}_{i}$ depends on the space dimension $d$, and
therefore on $\ep$.  Possibly, the simplest example along this line is
given by the field $\phi_{1}\equiv{\cal N}_{1}\varphi$, which we
describe in section \ref{sub:The-scaling-field-phi_1}.

Of course, before solving the ERG for the composite operators and
their products, we need to solve the ERG equation to obtain
$\Gamma_* [\varphi]$ at the fixed point.  In this work we use
perturbation theory to solve the ERG explicitly.  Since solving for
$\Gamma_* \left[\varphi\right]$ is not the main focus of this paper,
we give the first order derivation of $\Gamma_* [\varphi]$ in
Appendix \ref{App:The-fixed-point-EAA-1st-order}.  For the present
purposes, it suffices to say that the EAA parametrized by
$g$ is expressed as
\begin{align}
\Gamma (g) [\varphi] 
&= \frac{1}{2} \int_p \varphi (p) \varphi (-p) v^{(2)} (g; p) \notag\\
&\quad +
  \frac{1}{4!} \int_{p_1,\cdots,p_4} \prod_{i=1}^4 \varphi (p_i)\,
  (2\pi)^d \delta \left(\sum_{i=1}^4 p_i - p\right)\, v^{(4)} (g;
  p_1,\cdots,p_4)\,,\label{eq:EAA-order_g}
\end{align}
where
\begin{subequations}
\begin{align}
v^{(2)} (g; p) &= p^2 + g v_1^{(2)}\,,\\
v^{(4)} (g; p_1,\cdots,p_4) &= g + g^2 v_2^{(4)} (p_1,\cdots,p_4)\,.
\end{align}
\end{subequations}
$v_1^{(2)}$ is a constant given by (\ref{eq:appendix-v12}), and
the momentum dependent $v_2^{(4)}$ is given by (\ref{eq:appendix-v24}).
The Wilson-Fisher fixed point corresponds to
$g=g_* = \frac{(4\pi)^2}{3} \ep$ up to order $\ep$.  We introduce the
high momentum propagator and its derivative by
\begin{subequations}
\begin{align}
  h(p) &\equiv \frac{1}{p^2 + R(p)}\,,\\
  f(p) &\equiv (p \cdot \partial_p + 2) h(p) = \frac{(2 - p
         \cdot \partial_p) R(p)}{\left(p^2 + R(p)\right)^2}\,.
\end{align}
\end{subequations}
Both $R(p)$ and $f(p)$ decay rapidly at $p \gg 1$.
To second order in $g$ the beta function is given by
\begin{equation}
\beta (g) = \beta_2 \, g^2 = \left(- 3 \int_q f(q) h(q)\right) g^2\,,
\end{equation}
where
\begin{equation}
\int_q f(q) h(q) = \frac{1}{(4\pi)^2} + \mathrm{O} (\ep)\,.
\end{equation}
Using this result, calculated in Appendix \ref{App:asymptotic-behaviors},
we obtain $g_*$ given in (\ref{eq:fixed-point-value}).

\subsection{The scaling field
  $\phi_{1}$ \label{sub:The-scaling-field-phi_1}}

To first order in $g$, we obtain
\begin{equation}
  \Gamma^{(2)} (g; p,-p) = p^2 + g v_1^{(2)}
\end{equation}
where
\begin{equation}
  v_1^{(2)} = \frac{1}{2-\ep} \int_q f(q)
\end{equation}
is a constant.  As $p \to \infty$, we obtain
\begin{equation}
  \Gamma^{(2)} (g;p,-p) \overset{p \to \infty}{\longrightarrow} p^2\,.
\end{equation}
Hence, to first order in $g$, the two-point function is the same as
the Gaussian theory:
\begin{equation}
  \vev{\varphi (p) \varphi (q)} = \frac{1}{p^2} \, (2\pi)^d \delta (p+q)\,.
\end{equation}
Taking the inverse Fourier transform, we obtain
\begin{equation}
  \vev{\varphi (r) \varphi (0)} = \frac{1}{\N_1^2} \frac{1}{r^{d-2}}\,,
\end{equation}
where
\begin{equation}
  \frac{1}{\N_1^2} \equiv \frac{1}{4 \pi^{\frac{d}{2}}} \Gamma
  \left(\frac{d-2}{2}\right)\,.\label{eq:normalization-N1}
\end{equation}
(See Appendix \ref{App:Momentum-space-2pt-3pt}.)  Thus, $\phi_1 = \N_1
\varphi$ has the normalization of (\ref{eq:CFT-2pt-function-real-space}).

\subsection{The composite operator $\left[\varphi^{2}/2\right]$}

In this subsection we construct the composite operator
$\left[\varphi^{2}/2\right]$.  This satisfies the ERG equation
\begin{align}
&  \lb \left( \ep g + \beta (g) \right) \partial_g + p \cdot \partial_p
  + 2 + \Delta_\varphi \rb \op{\frac{1}{2} \varphi^2} (p) = \gamma_2
  (g) \op{\frac{1}{2} \varphi^2} (p) \notag\\
  &\quad + \int_q (2-q\cdot\partial_q) R(q) \frac{1}{2} \int_{r,s} G
    (g)_{q,-r} G(g)_{-s,-q} \frac{\delta^2}{\delta \varphi (r) \delta
    \varphi (s)} \op{\frac{1}{2} \varphi^2} (p)\,,
\end{align}
where $\gamma_2 (g)$ is the anomalous dimension.  To solve this, we
expand the operator in powers of $\varphi$:
\begin{equation}
  \op{\frac{1}{2} \varphi^2} (p) = \sum_{n=0}^\infty \frac{1}{(2n)!}
  \int_{p_1,\cdots,p_{2n}} \prod_{i=1}^{2n} \varphi (p_i)\cdot (2\pi)^d
  \delta \left(\sum_{i=1}^{2n} p_i - p \right)\, c^{(2n)} (g;
  p_1,\cdots,p_{2n})\,.
  \end{equation}
  We normalize the operator by the condition
\begin{equation}
  c^{(2)} (g;0,0) = 1\,.\label{eq:phi2-normalization}
\end{equation}
This determines $\gamma_2 (g)$.  To order $g$, we only need the first
two terms $n=0, 1$.  We expand
\begin{subequations}
  \begin{align}
    c^{(2)} (g;p_1,p_2) &= c^{(2)}_0 (p_1,p_2) + g\, c^{(2)}_1 (p_1,
                          p_2)\,,\\
    c^{(0)} (g) &= c^{(0)}_0 + g \,c^{(0)}_1\,\\
    \gamma_2 (g) &= g\, \gamma_{2,1}\,.
  \end{align}
\end{subequations}

\subsubsection{Order $g^{0}$}

(\ref{eq:phi2-normalization}) gives
\begin{equation}
  c^{(2)}_0 (p_1, p_2) = 1\,,
\end{equation}
independent of momenta.  For $n=0$, the ERG equation gives
\begin{equation}
c^{(0)}_0 = - \frac{1}{2-\ep} \frac{1}{2} \int_q f(q) = v^{(2)}_1\,.
\end{equation}

\subsubsection{Order $g^{1}$}

For $n=1$, the ERG equation gives
\begin{equation}
  \left( \sum_{i=1,2} p_i \cdot \partial_{p_i} + \ep \right) c^{(2)}_1 (p_1, p_2) =
  \gamma_{2, 1} - \int_q f(q) h(q+p)\,.
\end{equation}
(\ref{eq:phi2-normalization}) gives
\begin{equation}
  \gamma_{2,1} = \int_q f(q) h (q) = - \frac{1}{3} \beta_2 \overset{\ep \to
    0+}{\longrightarrow} \frac{1}{(4 \pi)^2}\,.
\end{equation}
Hence, the equation becomes
\begin{equation}
   \left( \sum_{i=1,2} p_i \cdot \partial_{p_i} + \ep \right)
   c^{(2)}_1 (p_1, p_2) = - 
   \int_q f(q) \left( h(q+p) - h (q) \right)\,.
 \end{equation}
This has no homogeneous solution analytic at $p_1=p_2=0$.  Hence, the
solution is given uniquely by
\begin{equation}
  c^{(2)}_1 (p_1, p_2) = - F(p)\,,
\end{equation}
where we define
\begin{equation}
  F (p) \equiv \frac{1}{2} \int_q h(q) \left(
    h(q+p)-h(q)\right)\,.\label{eq:F-definition} 
\end{equation}
For $n=0$, the ERG equation gives
\begin{equation}
  \left(- 2 + 2 \ep \right) c^{(0)}_1 = 0\,.
\end{equation}

We have thus obtained
\begin{align}
  \op{\frac{1}{2} \varphi^2} (p)
  &= \frac{1}{2} \int_{p_1, p_2} \varphi (p_1) \varphi (p_2) \delta
    (p_1+p_2-p) \lb 1 - g F(p)\rb\notag\\
  &\quad + v^{(2)}_1 \, (2\pi)^d \delta (p)\,.\label{eq:varphi2-expansion}
\end{align}
The anomalous dimension is
\begin{equation}
\gamma_2 (g_*) \simeq \gamma_{2,1} g_* = \frac{1}{3} \ep + \mathrm{O} (\ep^2)\,.
\end{equation}

\subsection{The scaling field $\phi_{2}$}

To determine the normalization
\begin{equation}
  \phi_2 (p) = \N_2 \op{\frac{1}{2} \varphi^2} (p)\,,
\end{equation}
we need to compute the two-point function of $\op{\frac{1}{2}
  \varphi^2}$.

Let
\begin{align}
  &  \op{\frac{1}{2} \varphi^2 \frac{1}{2} \varphi^2} (p, q) =
    \op{\frac{1}{2} \varphi^2} (p) \op{\frac{1}{2} \varphi^2} (q)
    \notag\\  & \quad +
  \int_{r,s} \frac{\delta}{\delta \varphi (r)}  \op{\frac{1}{2}
    \varphi^2} (p) \cdot G(g)_{-r,-s} [\varphi] \cdot
    \frac{\delta}{\delta \varphi (s)}\op{\frac{1}{2} \varphi^2} (q)
    + \QQ_{22} (g; p, q)\,.
\end{align}
The 1PI part is determined by the ERG equation
\begin{align}
  & \lb \left(\ep g + \beta (g)\right) \partial_g + p \cdot \partial_p
    + q \cdot \partial_q + 4 + \Delta_\varphi \rb
    \QQ_{22} (g; p,q) \notag\\
  &= 2 \gamma_2 (g) \QQ_{22} (g; p, q) +
    \gamma_{22,0} (g) (2\pi)^d \delta (p+q)\notag\\
  &\quad+ \int_{r} (2 - r \cdot \partial_r) R(r) \int_{s,t}
     G (g)_{r,s} G(g)_{t,-r} \left( \frac{1}{2} \frac{\delta^2}{\delta
    \varphi (-s) \delta \varphi (-t)} \QQ_{22} (g;p,q) \right.\notag\\
  &\qquad\left. +
    \int_{u,v} \frac{\delta^2}{\delta \varphi (-s) \delta \varphi (u)}
    \op{\frac{1}{2} \varphi^2} (p) G(g)_{-u, v} \frac{\delta^2}{\delta
    \varphi (-v) \delta \varphi (r)} \op{\frac{1}{2} \varphi^2 (q)}\right)\,.
\end{align}
We expand
\begin{equation}
  \QQ_{22} (g; p,q) = \sum_{n=0}^\infty \frac{1}{(2n)!} \int_{p_1,
    \cdots, p_{2n}} \prod_{i=1}^{2n} \varphi (p_i)\, (2\pi)^d \delta
  \left( \sum_{i=1}^{2n}p_i - p - q \right)\, d^{(2n)} (g; p,q;
  p_1,\cdots,p_{2n})\,.
\end{equation}
The mixing $\gamma_{22,0} (g)$ is determined by the normalization condition
\begin{equation}
  d^{(0)} (g; 0,0) = 0\,.\label{eq:normalization-Q22}
\end{equation}
To first order in $g$, only $d^{(0)}$ and $d^{(2)}$ are
non-vanishing.  We expand
\begin{subequations}
\begin{align}
d^{(2)} (g; p,q; p_1, p_2) &= g\, d_1^{(2)} (p,q; p_1, p_2)\,,\\
d^{(0)} (g; p,-p) &= d_0^{(0)} (p,-p) + g\, d_1^{(0)} (p,-p)\,,\\
\gamma_{22,1} (g) &= \gamma_{22,1;0} + g \,\gamma_{22,1;1}\,.
\end{align}
\end{subequations}

\subsubsection{Order $g^0$}

The ERG equation is
\begin{equation}
  \left( p \cdot \partial_p + \ep \right) d^{(0)}_0 (p,-p)
  = \gamma_{22,0;0} + \int_r f(r) h(r+p)\,.
\end{equation}
(\ref{eq:normalization-Q22}) gives
\begin{equation}
  \gamma_{22,0;0} = - \int_r f(r) h(r) = \frac{1}{3} \beta_2\,.
\end{equation}
Hence, we obtain
\begin{equation}
  d^{(0)}_0 (p,-p) = F (p)\,.
\end{equation}

\subsubsection{Order $g^1$}

For $n=1$, the ERG equation is
\begin{align}
&  \left( \sum_{i=1}^2 p_i \cdot \partial_{p_i} + p \cdot \partial_p +
  q \cdot \partial_q + 2 + \ep \right) d^{(2)}_1 (p,q; p_1,
  p_2)\\
  &= (-) \int_q \left( f(r) h(r+p) h(r+p+q) + h(r) f(r+p) h (r+p+q) +
    h(r) h(r+p) f(r+p+q) \right)\,.\notag
\end{align}
The solution, analytic at zero momenta, is given uniquely as
\begin{equation}
  d^{(2)}_1 (p,q; p_1, p_2) = (-) \int_r h(r) h(r+p) h(r+p+q)\,.
\end{equation}

For $n=0$, the ERG equation is
\begin{align}
  &\left( p \cdot \partial_p + 2 \ep \right) d^{(0)}_1 (p,-p) =
  \gamma_{22,0;1} - \frac{1}{2} \int_q f(q) \int_r h(r)^2
  h(r+p)\notag\\
  &\quad - v_1^{(2)} \int_r \left( 2 f(r) h(r) h(r+p) + h(r)^2 f(r+p)
    \right) - 2 \int_r f(r) h(r+p)\, F(p)\,.
\end{align}
(\ref{eq:normalization-Q22}) determines
\begin{equation}
  \gamma_{22,0;1} =  2 \ep v^{(2)}_1  \int_p h(p)^3\,.
\end{equation}
The solution is
\begin{equation}
 d^{(0)}_1 (p,-p) = - v_1^{(2)} \int_r h(r)^2 \left( h(r+p) -  h(r)\right) - F(p)^2\,.
\end{equation}

Thus, we obtain
\begin{equation}
  d^{(0)} (g_*; p,-p) = F(p) + g_* \lb - v^{(2)}_1 \int_r h(r)^2 \left(
    h(r+p) - h(r) \right) - F(p)^2 \rb\,.
\end{equation}
For large $p$, this gives the two-point function of
$\op{\frac{1}{2} \varphi^2}$.  The asymptotic behavior of $F(p)$ for large
momentum is obtained in (\ref{eq:Appendix-Fasymp}).  We then obtain the
two-point function as
\begin{align}
  \vev{\frac{\varphi^2}{2} (p) \frac{\varphi^2}{2} (q)}
  &= (2\pi)^d \delta (p+q)\, \left(F_{\mathrm{asymp}} (p) - g_*
    F_{\mathrm{asymp}} (p)^2 \right)\notag\\
  &= (2\pi)^d \delta (p+q)\, \frac{1}{2 (4\pi)^2} \left[ - \ln
    \frac{p^2}{4} + \mathrm{const} \right.\notag\\
&\left.\quad + \ep \lb \frac{1}{12} \left( \ln
    \frac{p^2}{4}\right)^2
 + \ln \frac{p^2}{4} \cdot \frac{1}{6} \left( \gamma - 2
  - \ln \pi + \frac{4}{3} (4 \pi)^2 \beta'_2 (0) \right) \rb \right]\,.
\label{eq:unnormalized-22}
\end{align}

\subsubsection{Normalization for $\phi_{2}$}

We wish to determine the constant ${\cal N}_{2}$ so that
$\phi_{2}={\cal N}_{2}\left[\varphi^{2}/2\right]$ is normalized as in
(\ref{eq:CFT-2pt-function-real-space}).  We expect $\N_2$ to differ
from the value at the Gaussian fixed point at order $\ep$, and we
parametrize it as
\begin{equation}
\frac{1}{\N_2^2} = \frac{1}{2} \left( \frac{1}{4 \pi^{\frac{d}{2}}} \Gamma
  \left(\frac{d-2}{2}\right) \right)^2 \left(1 + A\, \ep \right)\,.\label{eq:normalization-N2}
\end{equation}

Since the scale dimension of $\phi_2$ (in coordinate space) is 
\begin{equation}
\Delta_2 = d-2+\gamma_2 (g_*) \simeq 2 - \frac{2}{3} \ep \label{eq:Delta2-epsilon}
\end{equation}
we expect, from (\ref{eq:CFT_2pt-function_momentum_space}),
\begin{align}
\vev{\frac{1}{2} \varphi^2 (p) \frac{1}{2} \varphi^2 (q)}
&= \frac{1}{\N_2^2} \vev{\phi_2 (p) \phi_2 (q)}\notag\\
&= (2 \pi)^d \delta (p+q) \, \frac{1}{\N_2^2} \pi^{\frac{d}{2}}
  \frac{\Gamma \left(\frac{d}{2}- 4 + \frac{4}{3} \ep\right)}{\Gamma
  \left(2 - \frac{2}{3} \ep\right)}
  \left(\frac{4}{p^2}\right)^{\frac{d}{2}-2 + \frac{2}{3} \ep}\notag\\
&=  (2 \pi)^d \delta (p+q) \, \frac{1}{32} \pi^{-2+\frac{\ep}{2}}
  \frac{\Gamma \left(1-\frac{\ep}{2}\right)^2 \Gamma
  \left(\frac{\ep}{6}\right)}{\Gamma \left(2 - \frac{2}{3} \ep\right)}
  \left(\frac{4}{p^2}\right)^{\frac{\ep}{6}} \cdot (1 + A\,
  \ep)\notag\\
&\simeq (2 \pi)^d \delta (p+q) \, \frac{1}{2 (4 \pi)^2} \left[  - \ln
  \frac{p^2}{4} + \mathrm{const} \right.\notag\\
&\left.\quad + \ep  \lb \frac{1}{12} \left( \ln
  \frac{p^2}{4}\right)^2 +  \ln \frac{p^2}{4} \cdot \frac{1}{6}
  \left(- 6 A - 4 - \gamma - 3  \ln \pi \right)  \rb \right]
\end{align}
Comparing this with (\ref{eq:unnormalized-22}), we obtain
\begin{equation}
A = - \frac{1}{3} \left(1 + \gamma + \ln \pi + \frac{2}{3} (4 \pi)^2
  \beta'_2 (0) \right)\,.\label{eq:result-A}
\end{equation}

\subsection{The composite operators $\left[\varphi^{4}/4!\right]$ and $\phi_4$}

In this subsection we construct the composite operator
$\op{\varphi^4/4!}$.  This satisfies the ERG equation
\begin{align}
&\lb \left( \ep g + \beta (g) \right) \partial_g + p \cdot \partial_p +
  \ep + \Delta_\varphi \rb \op{\frac{1}{4!} \varphi^4 } (p) = \gamma_4
  (g) \op{\frac{1}{4!} \varphi^4} (p) \notag\\
& + \gamma_{4,\phi \partial^2\phi} \op{\varphi \partial^2 \varphi } (p) +
  \gamma_{4, \partial^2 \varphi^2/2} (-p^2) \op{\frac{1}{2}
  \varphi^2} (p)\notag\\
& + \int_q (2 - q \cdot \partial_q) R(q) \, \frac{1}{2} \int_{r,s}
  G(g)_{q,-r} G(g)_{-s,-q} \frac{\delta^2}{\delta \varphi (r) \delta
  \varphi (s)} \op{\frac{1}{4!} \varphi^4} (p)\,,
\end{align}
where $\gamma_4 (g)$ is the anomalous dimension.  To solve this, we
expand
\begin{equation}
\op{\frac{1}{4!} \varphi^4} (p) = \sum_{n=0}^\infty \frac{1}{(2n)!}
  \int_{p_1,\cdots,p_{2n}} \prod_{i=1}^{2n} \varphi (p_i) \cdot
  (2\pi)^d \delta \left(\sum_{i=1}^{2n} p_i - p \right) c^{(2n)}
  (g;p)_1,\cdots, p_{2n})\,.
\end{equation}
We normalize the operator by the condition
\begin{equation}
c^{(4)} (g; 0,0,0,0) = 1\,.\label{eq:varphi4-normalization}
\end{equation}
This determines $\gamma_4 (g)$.  The mixing is determined by the
additional boundary conditions at $p_1=p_2=0$:
\begin{equation}
c_2 (g;p_1, p_2) - c_2 (g; 0,0) = \mathrm{O} (p^4)\,.\label{eq:varphi4-boundaryconditions}
\end{equation}
To order $g$, we only need the first
three terms $n=0,1,2$.  We expand
\begin{subequations}
\begin{align}
c^{(4)} (g;p_1,\cdots,p_4) &= c^{(4)}_0 + g \,c^{(4)}_1
                             (p_1,\cdots,p_4)\,,\\
c^{(2)} (g; p_1, p_2) &= c^{(2)}_0 + g \,c^{(2)}_1 (p_1, p_2)\,,\\
c^{(0)} (g) &= c^{(0)}_0 + g \,c^{(0)}_1\,,\\
\gamma_4 (g) &= g\, \gamma_{4,1}\,.
\end{align}
\end{subequations}
We do not need to calculate $c^{(0)}$.

\subsubsection{Order $g^0$}

For $n=2$, (\ref{eq:varphi4-normalization}) gives
\begin{equation}
c^{(4)}_0 = 1\,.
\end{equation}
For $n=1$, the ERG equation gives
\begin{equation}
c^{(2)}_0 = - \frac{1}{2-\ep} \frac{1}{2} \int_q f(q) =  v^{(2)}_1\,.
\end{equation}

\subsubsection{Order $g^1$}

For $n=2$, the ERG equation gives
\begin{align}
& \left( \sum_{i=1}^4 p_i \cdot \partial_{p_i} + \ep \right) c^{(4)}_1
  (p_1,\cdots,p_4)\notag\\
&= \gamma_{4,1} - \int_q f(q) \left( h(q+p_1+p_2) + h(q+p_3+p_4) +
  (\textrm{t-, u-channels}) \right)\,.
\end{align}
The normalization condition (\ref{eq:varphi4-normalization}) at $p_i=0$
requires
\begin{equation}
\gamma_{4,1} = 6 \int_q f(q) h(q) = - 2 \beta_2\,.
\end{equation}
We then obtain
\begin{equation}
c^{(4)}_1 (p_1,\cdots,p_4) = - F (p_1+p_2) - F(p_3+p_4) -  (\textrm{t-,
  u-channels})\,.
\end{equation}

For $n=1$, the ERG equation gives
\begin{align}
& \left(\sum_{i=1}^2 p_i \cdot \partial_{p_i} + 2 \ep - 2 \right)
  c^{(2)}_1 (p_1,p_2)\notag\\
&= - 2 \beta_2 v^{(2)}_1 + \gamma_{4,\varphi\partial^2\varphi} (-)
  (p_1^2 + p_2^2) + \gamma_{4,\partial^2 \varphi^2/2} (-p^2)\notag\\
&\quad - v^{(2)}_1 \int_q f(q) h(q+p)  - \frac{1}{2} \int_q f(q)
  \cdot F(p)\notag\\
&\quad - v^{(2)}_1 \int_q f(q) h(q) - \int_q f(q) \left( F(q+p_1) +
  F(q+p_2)\right)\,.
\end{align}
The boundary conditions (\ref{eq:varphi4-boundaryconditions}) at
$p_1=p_2=0$ demand
\begin{subequations}
\begin{align}
\gamma_{4,\varphi\partial^2\varphi} &= \eta_2\,,\\
\gamma_{4,\partial^2\varphi^2/2} &= - 2 \ep v^{(2)}_1 \frac{d}{dp^2}
                                   F(p=0)\,,
\end{align}
\end{subequations}
where
\begin{equation}
\eta_2 \equiv - \frac{d}{dp^2} \int_q f(q) F(q+p)\Big|_{p=0}\,.
\end{equation}
The solution is given by
\begin{equation}
c^{(2)}_1 (p_1, p_2) = - v^{(2)}_1 \left( F (p) - p^2 \frac{d}{dp^2}
                       F(p=0) \right) - G(p_1) - G(p_2)\,.
\end{equation}
The function $G(p)$ is defined by
\begin{equation}
\left( p \cdot \partial_p  - 2 + 2 \ep \right) G(p)
= \int_q f(q) F(q+p) + v^{(2)}_1 \frac{2}{3} \beta_2 + \eta_2 \,p^2\,,
\end{equation}
and the condition
\begin{equation}
\frac{d}{dp^2} G(p)\Big|_{p=0} = 0\,.
\end{equation}

We have thus obtained
\begin{align}
\op{\frac{1}{4!} \varphi^4} (p)
&= \frac{1}{4!} \int_{p_1,\cdots,p_4} \prod_{i=1}^4 \varphi (p_i)\,
  (2\pi)^d \delta \left( \sum_{i=1}^4 p_i - p \right)\, \left( 1 
- g \left( F(p_1+p_2) + F(p_3+p_4) + \cdots
\right)  \right)\notag\\
&\quad + \frac{1}{2} \int_{p_1,p_2} \varphi (p_1) \varphi (p_2)
  \,(2\pi)^d \delta (p_1+p_2-p)\notag\\
&\qquad \times  \lb v^{(2)}_1 \left( 1 -  g
  \left( F(p) - p^2 \frac{d}{dp^2} F(0) \right)  \right) - g \left(
  G(p_1) + G(p_2) \right) \rb\,.\label{eq:varphi4-expansion}
\end{align}

\subsubsection{The normalization ${\cal N}_{4}$}

Before discussing the normalization of the scaling operator $\phi_4$,
let us recall that $\phi_4$ is actually a linear combination of
$\left[\varphi^4/4! \right]\left(x\right)$,
$\varphi \partial^2 \varphi$, and the total derivative
$\partial^2 \varphi^2/2$.  The operator $\varphi \partial^2 \varphi$
is an equation of motion operator whose correlation functions vanish
at separate points (the equation of motion operator contributes to
correlation functions only via delta functions).  Hence, the mixing
with $\varphi \partial^2 \varphi$ is totally harmless.  
Thus, the mixing with the total derivative and the equation of motion operator can be neglected for our present purposes.\footnote{
Note that, since we are interested at the OPE coefficient at the Wilson-Fisher fixed point, we could solve directly the fixed point equation for the scaling composite operators and their products. This leads to the same final OPE coefficient but the calculations are somewhat more cumbersome.}

We determine the normalization factor ${\cal N}_{4}$ to the leading
order in $\ep$, i.e., O$\left(\ep^{0}\right)$.  Since
$c_{114}$
vanishes in the Gaussian theory, we have $c_{114}\sim\ep$,
and the O$\left(\ep\right)$
correction to $\N_{4}$ is relevant only for the OPE coefficient to higher
orders. Therefore, we can take the Gaussian value of $\N_4$ given in
(\ref{eq:appendix-Gauss-N4}):
\begin{equation}
\frac{1}{\N_4^2} = \frac{1}{4!} \frac{1}{\N_1^8} = \frac{1}{4! 2^8 \pi^{2d}}
\Gamma \left(\frac{d-2}{2}\right)^4 \,.\label{eq:normalization-N4}
\end{equation}
It is also possible to obtain this result directly in the ERG
framework.  The computation in momentum space requires three-loop
calculations, and we do not display it here.

\section{OPE coefficients from
  ERG \label{sec:OPE-coefficients-from-ERG}}

As mentioned in section
\ref{sub:Operator-product-expansion-coeff-in-the-ERG} we will deduce
the OPE coefficients from the three-point functions of the associated
operators. In this section we compute the coefficients $c_{112}$ and
$c_{114}$ to order $\ep$ at the Wilson-Fisher fixed point. The
three-point functions involved are respectively
$\langle\phi_{1}\phi_{1}\phi_{2}\rangle$ and
$\langle\phi_{1}\phi_{1}\phi_{4}\rangle$.  We have already explained
how to compute these within the EAA formalism in section 
\ref{sec:Wilson-Fisher-fp}.

Using (\ref{eq:formula-three-point}), we obtain
\begin{subequations}
\label{eq:normalized-3pt-function-via-EAA}
\begin{align}
\frac{1}{\N_1^2 \N_2}
  \vev{\phi_{1}\left(-p_{1}\right)\phi_{1}\left(-p_{2}\right)\phi_{2}\left(p\right)} 
   &= \Bigr\langle\varphi\left(-p_{1}\right) \varphi
     \left(-p_{2}\right)\left[\frac{\varphi^{2}}{2}\right]\left(p\right)\Bigr\rangle\,.   
     \notag\\
&=  \frac{1}{p_1^2 p_2^2} \op{\frac{1}{2} \varphi^2}^{(2)} (p_1, p_2)
 \, (2\pi)^d \delta (p_1+p_2-p)\,,\\
\frac{1}{\N_1^2 \N_4} \vev{\phi_1 (-p_1) \phi_1 (-p_2) \phi_4 (p)}
&= \vev{\varphi (-p_1) \varphi (-p_2) \op{\frac{1}{4!}
  \varphi^4} (p)}\notag\\
&= \frac{1}{p_1^2 p_2^2}\, \op{\frac{1}{4!}
  \varphi^4}^{(2)} (p_1, p_2)\, (2\pi)^d \delta (p_1+p_2-p)\,,
\end{align}
\end{subequations}
where we have to take the momenta much larger than the fixed cutoff of
order $1$.

We now have all the ingredients to compute the three-point functions
of the normalized fields.  In section
\ref{sec:Scaling-operators-from-ERG} we have determined the
normalization constants ${\cal N}_{i}$ and the composite operator
vertex functions $\op{\Op}^{(2)}$ 
entering (\ref{eq:normalized-3pt-function-via-EAA}).

To finally read off the OPE coefficient we compare the limit $p_{1}\gg p_{2}$
of (\ref{eq:normalized-3pt-function-via-EAA}) with the expectation
from CFT (\ref{eq:CFT_3pt-function_momentum_space}) with an unknown
OPE coefficient $c_{11i}$. It is interesting
to note that taking either one of $p_{1}$ or $p_{2}$ much larger
than the other corresponds to taking an OPE in the three-point function.
For instance, if we take $p_{1}\gg p_{2}$ we are effectively taking
the short distance limit of the product
$\phi_{1}\left(-p_{1}\right)\phi_{i}\left(p_{1}+p_{2}\right)$. 
Of course, if one applies the same reasoning to the case $p_{2}\gg p_{1}$
one can read off the same OPE coefficient. This is manifest in our
vertices since they are symmetric under $p_{1}\leftrightarrow p_{2}$.
Taking one of the momenta much larger than the other corresponds
to a short distance limit, and the comparison of the two limits $p_{1}\gg p_{2}$
and $p_{2}\gg p_{1}$ reminds us of the bootstrap associativity requirements.

\subsection{The OPE coefficient $c_{112}$} \label{subsec:c_112}

We expand the OPE coefficient
\begin{equation}
c_{112} = \sqrt{2} \left( 1 + B \, \ep \right) \label{eq:c112-expansion}
\end{equation}
to first order in $\ep$, where $\sqrt{2}$ is the value for the
Gaussian theory.  We compute $B$ by comparing the result obtained from
the ERG with that expected from CFT.  For completeness, we also sketch
an alternative method by constructing the operator product
$\left[\phi_{1}\phi_{2}\right]$ and expanding it up to the operator
$\phi_{1}$.

\subsubsection{From the three point function}

Using a result (\ref{eq:CFT_3pt-function_momentum_space}) from CFT, we
obtain
\begin{align}
\vev{\phi_1 (-p_1) \phi_1 (-p_2) \phi_2 (p)} &\overset{p_1 \gg p_2}{\longrightarrow} (2
  \pi)^d \delta (p-p_1-p_2) \frac{c_{112}}{p_1^{d - \Delta_2}
  p_2^2}\notag\\
&\quad \times \frac{(4 \pi)^d}{4^{d-2+\frac{1}{2} \Delta_2}}
  \frac{\Gamma \left( \frac{1}{2} (d-\Delta_2)\right)}{\Gamma \left( \frac{1}{2}
  \Delta_2\right) \Gamma \left(\frac{1}{2}(d-2)\right)}\,,
\end{align}
where we have substituted the Gaussian value $\Delta_1 = \frac{1}{2}
(d-2)$.  Now, using the results from section
\ref{sec:Scaling-operators-from-ERG}, i.e., $\Delta_2$ (\ref{eq:Delta2-epsilon}),
$\N_1$ (\ref{eq:normalization-N1}),  and $\N_2$
(\ref{eq:normalization-N2}, \ref{eq:result-A}),
we obtain
\begin{align}
  &\vev{\varphi (-p_1) \varphi (-p_2) \op{\frac{1}{2} \varphi^2} (p)}
    = \frac{1}{\N_1^2 \N_2} \vev{\phi_1 (-p_1) \phi_1 (-p_2) \phi_2
    (p)}\notag\\
  &\overset{p_1 \gg p_2}{\longrightarrow} \left(1 + B \ep \right)
    \left(1 + \frac{1}{2} A \ep\right) 
    \frac{\Gamma \left(1 - \frac{\ep}{2}\right) \Gamma \left(1 -
    \frac{\ep}{6}\right)}{\Gamma \left(1 - \frac{\ep}{3}\right)}
    \frac{1}{p_1^2 p_2^2} \left(
    \frac{p_1^2}{4}\right)^{\frac{\ep}{6}}\notag\\
  &\quad = \frac{1}{p_1^2 p_2^2} \left[ 1 + \epsilon \left\lbrace B
+ \frac{1}{6} \left( \ln \frac{p_1^2}{4} + \gamma - \ln \pi - 1 -
  \frac{2}{3} (4 \pi)^2 \beta'_2 (0) \right) \right\rbrace \right]\,.
\end{align}
(We have omitted $(2\pi)^d \delta (p_1+p_2-p)$ for simplicity.)

We compare this with the result from (\ref{eq:varphi2-expansion}):
\begin{align}
\op{\frac{1}{2} \varphi^2}^{(2)} (p_1, p_2)
&\overset{p_1 \to \infty}{\longrightarrow}
1 - g_* F_{\mathrm{asymp}} (p_1)\notag\\
&= 1 + \frac{\ep}{6} \left[ \ln \frac{p_1^2}{4} + \gamma - \ln \pi  - 2 -
  \frac{2}{3} (4\pi)^2 \beta_2' (0) \right]\,.
\end{align}
Hence, we obtain
\begin{equation}
B = - \frac{1}{6}\,.
\end{equation}
In conclusion we have obtained
\begin{equation}
c_{112} = \sqrt{2} \left( 1 - \frac{1}{6} \ep \right)
\label{eq:result-c_112}
\end{equation}
to order $\ep$, in agreement with the results in the literature \cite{Gopakumar:2018xqi}.

Finally, it is interesting to note that the final result (\ref{eq:result-c_112}) is
independent of the form of the cutoff kernel $R(p)$. This signals the
physical nature of the OPE coefficients.
In the present case, this is achieved by a cancellation of cutoff dependent factors present in ${\cal N}_2$ and $\left[\frac{\varphi^2}{2} \right]^{\left(2\right)}$.

\subsubsection{From the operator product}

We consider the operator product
$\left[\varphi\left[\varphi^{2}/2\right]\right]$ and expand it up to
a term linear in the fields:
\begin{equation}
\left[\varphi\left(p_{1}\right)\left[\frac{\varphi^{2}}{2}\right]\left(p_{2}\right)\right]  
=  \frac{1}{\N_2} C_{121}
\left(\frac{p_{1}-p_{2}}{2}\right) \varphi \left(p_{1}+p_{2}\right)+\cdots\,.
\end{equation}

Let us first consider what is expected for the Wilson coefficient in
momentum space. We recall that $C_{121} \sim x^{-\Delta_{2}}$, and
$\Delta_{2}\simeq 2-\frac{2}{3}\ep$. To linear order in $\ep$
one has
\begin{equation}
C_{121} (P)
= \sqrt{2} \, \frac{4 \pi^2}{P^2} \left[ 1+ \ep \left( B -
  \frac{\gamma}{6} + \frac{1}{6} \ln P^2 - \frac{1}{2} \ln \pi - \frac{\ln
  2}{3} \right) \right]\,.
\end{equation}

On the other hand, ERG gives
\begin{align}
\op{\varphi (p) \op{\frac{1}{2} \varphi^2} (q)}
&\overset{p \to \infty}{\longrightarrow} \frac{1}{p^2} \op{\frac{1}{2}
  \varphi^2}^{(2)} (-p, q+p) \varphi (q+p)\notag\\
&= \frac{1}{p^2} \left(1 - g_* F_{\mathrm{asymp}} (q)\right) \varphi (q+p)\,.
\end{align}
The comparison reproduces
\[
c_{112} \simeq \sqrt{2}\left(1-\frac{\ep}{6}\right)\,.
\]

\subsection{The OPE coefficient $c_{114}$} \label{subsec:c_114}

At order $O\left(g^{0}\right)$ the theory is Gaussian so that
$c_{114} = 0$ to order $\ep^0$.  We then expect
\begin{equation}
c_{114}= C \ep + \mathrm{O} (\ep^2)\,.
\end{equation}
Hence, we can use the Gaussian values for the scale dimension 
\begin{equation}
  \Delta_4 = 2 (d-2) = 4 - 2 \ep
\end{equation}
and $\N_4$ (\ref{eq:normalization-N4}).

Hence,  from (\ref{eq:CFT_3pt-function_momentum_space}), we expect
\begin{align}
&\vev{\varphi (-p_1) \varphi (-p_2) \op{\frac{1}{4!} \varphi^4} (p)}
= \frac{1}{\N_1^2 \N_4} \vev{\phi_1 (-p_1) \phi_1 (-p_2) \phi_4
  (p)}\notag\\
&\overset{p_1 \to \infty}{\longrightarrow} (2 \pi)^d \delta
(p_1+p_2-p) \frac{c_{114}}{p_1^\ep p_2^2} 
\,  \frac{1}{\sqrt{4!}} \frac{1}{(4 \pi)^{2-\frac{\ep}{2}}}
\Gamma \left(1 - \frac{\ep}{2}\right)^2 \frac{\Gamma
  \left(\frac{\ep}{2}\right)}{\Gamma (2-\ep)}\notag\\
&\qquad =  (2 \pi)^d \delta (p_1+p_2-p) \,\frac{1}{p_1^2 p_2^2}\, C \ep\, \frac{1}{\sqrt{4!}
  (4 \pi)^2} p_1^2 \left( - \ln \frac{p_1^2}{4} +
  \mathrm{const} \right)\,.
\end{align}

We compare this with the result from (\ref{eq:varphi4-expansion}):
\begin{align}
\op{\frac{1}{4!} \varphi^4}^{(2)} (p_1, p_2) &\overset{p_1 \to
                                               \infty}{\longrightarrow}
- g_* G_{\mathrm{asymp}} (p_1)\notag\\
&= \ep \, \frac{(4 \pi)^2}{3} \frac{1}{2 \cdot 3!}\frac{1}{(4\pi)^4}
  p_1^2 \left( - \ln p_1^2 + \mathrm{const} \right)\,,
\end{align}
where we have used the asymptotic form $G_{\mathrm{asymp}} (p)$
obtained in (\ref{eq:Appendix-Gasymp}).  Hence, we obtain
\begin{equation}
C = \frac{\sqrt{4!}}{6 \cdot 3!} = \frac{1}{\sqrt{54}}\,.
\end{equation}

In conclusion we have obtained
\begin{equation}
c_{114}  =  \frac{\ep}{\sqrt{54}}
\end{equation}
in agreement with the results in the literature \cite{Gopakumar:2018xqi}.

\subsection{Extension to other systems}

Let us mention that the strategy developed in this work is rather
general, and it applies to a wide variety of models. For instance,
unitarity is not an essential ingredient, and we can as well compute
the OPE coefficients of non-unitary theories. In order to see this
explicitly, in this subsection we compute the OPE coefficient $c_{111}$
in the Lee-Yang model in $d = 6-\ep$ dimensions.

The Lee-Yang model has been studied via ERG even beyond the
perturbative regime \cite{An:2016lni,Zambelli:2016cbw}. Here, we limit
ourselves to computing the leading perturbative correction to
$c_{111}$, which vanishes in the non-interacting theory. Let us
consider the action \cite{Fisher:1978pf}
\begin{equation}
S\left[\chi\right]  =  \int\left\{ \frac{1}{2}\partial_{\mu}
                         \chi \partial_{\mu} \chi + i\frac{g}{3!} \chi^{3}\right\} \,.
\end{equation}
In the one-loop approximation, it is possible to compute the wave
function renormalization and the beta function of the coupling $g$.
We do not reproduce these computations here, and merely report the
fixed-point value of the coupling, i.e.,
$g_* \simeq 8\sqrt{\frac{2\pi^{3}}{3}}\sqrt{\ep}$.  The coefficient is
cutoff independent.

We read off the OPE coefficient from the three-point function
$\langle\phi_{1}\phi_{1}\phi_{1}\rangle$.  In general one has
\begin{align}
&\langle\chi\left(x_{1}\right)\chi\left(x_{2}\right)\Op\left(x\right)\rangle_{{\rm conn}} 
 =  \int_{z_1,z_2} G\left(x_{1},z_{1}\right) 
\left(\frac{\delta\Gamma}{\delta\vep\left(x\right)
  \delta\varphi\left(z_{1}\right)
  \delta\varphi\left(z_{2}\right)}\right) G\left(z_{2},x_{1}\right) \notag\\
&\quad  -\int_{z_1,z_2,u_1,u_2} \frac{\delta^{2}\Gamma}{\delta \vep
  \left(x\right) \delta \varphi \left({z_{1}}\right)}
G\left(z_{1},z_{2}\right) \frac{\delta^3 \Gamma}{\delta \varphi
  \left(z_{2}\right) \delta\varphi\left(u_{1}\right)\delta\varphi\left(u_{2}\right)}
G\left(u_{1},x_{1}\right) G\left(u_{1},x_{2}\right) \,.
\label{eq:3-pt-function-general}
\end{align}
To the first order in $g_{*}$, we have
\begin{align}
\langle\phi_{1}\phi_{1}\phi_{1}\rangle 
& =  - i g_* {\cal N}_{1}^{3} G\left(p_{1}\right) G\left(p_{2}\right)
  G\left(p_{1}+p_{2}\right)\notag\\ 
 & \approx  -i\sqrt{\frac{2}{3}} \sqrt{\ep} \left(64\pi^{6}\right)
   \frac{1}{p_{1}^{4}p_{2}^{2}}\,, 
\end{align}
where ${\cal N}_{1}=\sqrt{4\pi^{3}}$ is the normalization of the
field. A straightforward comparison with the expectation from CFT gives
\begin{equation}
c_{111}  \simeq  -i\sqrt{\frac{2}{3}}\sqrt{\ep}\,,
\end{equation}
which agrees with other results in the literature
\cite{Gopakumar:2016cpb,Hasegawa:2016piv}.  Note that the imaginary
factor in $c_{111}$ is a clear sign of the non-unitarity of the
model. This does not prevent us from using our framework to compute
the OPE coefficients.

Let us note that in the present computation the anomalous dimension of
the operators involved was not necessary, only the
$O\left(\epsilon^0 \right)$-scaling dimensions enter the calculation.
This is because the OPE coefficient is trivial at leading order,
i.e., at order $O\left(\epsilon^0 \right)$.  As a consequence the
$O\left(\sqrt{\epsilon} \right)$-coefficient is determined by the
leading order composite operator and the EAA at order
$O\left(\sqrt{\epsilon} \right)$.  Generically, however, in order to
compute an OPE coefficient to a certain order it is necessary to
compute the anomalous dimension of all the operators involved to the
same order. Indeed the anomalous dimensions enter the
determination of the normalizations ${\cal N}_i$ and in the
construction of the relevant 1PI composite operator vertices, as in
the examples of sections \ref{subsec:c_112} and \ref{subsec:c_114}.

\subsection{On non-perturbative approximation schemes \label{sub:On-non-pert-approximatiox-schemes}}

In this work we solved the ERG equations perturbatively. However, ERG
is known to provide a framework for non-perturbative approximation
schemes, which allows one to provide also precise results.  See,
e.g., \cite{Canet:2003qd,Balog:2019rrg}.  It is natural then to ask if
any of such approximation schemes can be naturally employed within the
strategy outlined in this work.

The most widely employed approximation schemes are the derivative
expansion and the BMW scheme.\cite{Blaizot:2005wd} In the derivative
expansion one retains the full field dependence of the operators
appearing in the EAA up to a certain number of derivatives:
$\Gamma_{k}\left[\varphi\right]=\int V\left(\varphi\right)+\int
K\left(\varphi\right)\partial\varphi\partial\varphi+\cdots$.
In the BMW scheme, instead, one does not retain such a general field
dependence but aims to keep full momentum dependence with the use of
background fields.  We refer to \cite{Blaizot:2005wd} for a detailed
presentation.

The strategy outlined in section
\ref{sub:Operator-product-expansion-coeff-in-the-ERG} clearly relies
on having control over the momentum dependence of the composite
operator vertices. It follows that a derivative expansion type of
scheme is not suitable for our purposes. An ideal scheme may be the
BMW since it retains the momentum dependence.  Recently, the BMW
has been applied to compute the two-point function of the operator
$\varphi^{2}/2$ in relation to the ``Higgs amplitude mode''
\cite{Rose:2015bma}.  Such a study shows that it is indeed possible to
keep track of the momentum dependence of the composite operator
vertices in non-perturbative approximation schemes.  It may thus be
possible to apply our strategy even beyond perturbation theory.  
Note that a crucial ingredient in the BMW scheme is the use
of large external, i.e., non-loop, momenta. This fits well with the
recipe of looking at large momenta in order to read off the OPE
coefficients.

Finally let us mention a different approach studied in the literature.
Cardy proposed a formula that relates the second order expansion of
the beta function around a fixed point with the OPE coefficients
\cite{cardy1996scaling}.  Such a formula has been employed to obtain
leading order corrections to the OPE coefficients in the
$\ep$-expansions \cite{Codello:2017hhh}.  In its original formulation,
however, the proposed relation displays scheme dependent OPE
coefficients. It is possible, however, to define scheme independent
coefficients in the expansions of the beta functions around a fixed
point by introducing a connection on the theory space
\cite{Pagani:2017gnd}.  This may lead to a possible further strategy
that tackles the computation of the OPE coefficients via a geometric
viewpoint of the RG flow. However, it must be emphasized that,
strictly speaking, the argument by Cardy applies only to the coefficients
related to non-integrable singularities. For this reason, the strategy
outlined in this paper gives access to even less singular OPE coefficients.

\section{Summary and outlook } \label{sec:summary-and-outlook}

In this work we studied OPE within the ERG formalism and showed by
explicit computation that ERG can be employed to compute the OPE
coefficients.  Such OPE coefficients are independent of the RG
scheme employed once one fixes a normalization convention for the
operator content of the theory.

In section \ref{sec:Operator-product-expansion-in-the-ERG} we
introduced the ERG framework for composite operators and outlined our
strategy for the computation of the OPE coefficients. In section
\ref{sec:Wilson-Fisher-fp} we introduced a version of the ERG
framework suitable for discussions of a fixed point.  In section
\ref{sec:Scaling-operators-from-ERG} we studied explicitly some
examples of composite operators, for which we computed OPE
coefficients within the $\ep$-expansion in section
\ref{sec:OPE-coefficients-from-ERG}. Interestingly, as mentioned in
section \ref{sub:On-non-pert-approximatiox-schemes}, our strategy does
not rely on the use of perturbation theory. Perturbation theory has
been used as a way to provide an actual approximate solution to the
ERG equations of interest. However, by employing non-perturbative
approximation schemes it may be possible to go beyond perturbative
results, and we hope our work paves the way in this direction.

It must be said that the strategy employed in this work is not
expected to be as efficient as other methods in the literature to compute
the CFT data, in particular with respect to the bootstrap approach.
However, besides its conceptual relevance, the ERG framework allows
one to easily extend the methodology studied in this work to very
different systems, possibly even in the absence of conformal symmetry
or out of equilibrium. For this reason we think it worthwhile studying
OPE further within the ERG formalism.

\subsection*{Acknowledgments}

This work was supported by the DFG grant PA 3040/3-1.  C.P.~and
H.S.~would like to thank Kobe University and Johaness Gutenberg
Universit{\"a}t, respectively, for hospitality while part of this work
was done.

\newpage{}

\appendix

\section{Two- and three-point functions in momentum space\label{App:Momentum-space-2pt-3pt}}

In this section we derive the most important formulas used in the
main text. We adopt the following notation:
\[
\int_{x} \equiv \int d^{d}x\,,\quad
\int_{p} \equiv \int\frac{d^{d}p}{\left(2\pi\right)^{d}}\,.
\]
We denote by $\tilde{g} (p)$ the Fourier transform of $g (x)$:
\[
\tilde{g}\left(p\right)  \equiv
\int_{x}e^{-ipx}g\left(x\right)\quad\textrm{so that}\quad
g\left(x\right) = \int_{p}e^{ipx}\tilde{g}\left(p\right)\,.
\]

\subsection{Two-point functions}

The Fourier transform of the two-point function
\begin{equation}
f_2 (x) \equiv \frac{1}{x^{2\alpha}}
\end{equation}
is obtained as
\begin{equation}
\tilde{f}_2 (p) \equiv \int_{x}e^{ipx}\frac{1}{x^{2\alpha}} = \pi^{d/2} \frac{\Gamma
  \left(\frac{d}{2}-\alpha\right)}{\Gamma (\alpha)}
\left(\frac{4}{p^{2}}\right)^{\frac{d}{2}- \alpha}\,.
\label{eq:Four-of-normalized-2pt-function}
\end{equation}

Using the above results we can calculate the normalization constants
$\N_1$, $\N_2$, and $\N_4$ at the Gaussian fixed point.  The standard
normalization of $\varphi$ is given by the propagator in the momentum
space:
\begin{equation}
\vev{\varphi (p) \varphi (q)}_G = \frac{1}{p^2}\, (2\pi)^d \delta
(p+q)\,.
\end{equation}
This corresponds to $\alpha = \frac{d-2}{2}$ above.  Hence, the propagator in
the coordinate space is given by
\begin{equation}
\vev{\varphi (r) \varphi (0)}_G = \frac{1}{\N_1^2}
\frac{1}{r^{d-2}}\,,\label{eq:appendix-Gauss-N1} 
\end{equation}
where
\begin{equation}
\frac{1}{\N_1^2} = \frac{1}{4 \pi^{\frac{d}{2}}} \Gamma \left(\frac{d-2}{2}\right)\,.
\end{equation}
(\ref{eq:appendix-Gauss-N1}) implies
\begin{equation}
\vev{\frac{1}{2} \varphi^2 (r) \frac{1}{2} \varphi^2 (0)}_G =
\frac{1}{2} \vev{\varphi (r)\varphi (0)}_G^2 = \frac{1}{2 \N_1^4}
\frac{1}{r^{2(d-2)}}\,.
\end{equation}
Hence, we obtain
\begin{equation}
\frac{1}{\N_2^2} = \frac{1}{2 \N_1^4} = \frac{1}{32 \pi^d} \Gamma
\left(\frac{d-2}{2}\right)^2\,.\label{eq:appendix-Gauss-N2}
\end{equation}
Similarly,
\begin{equation}
\vev{\frac{1}{4!} \varphi^4 (r) \frac{1}{4!} \varphi^4 (0)}_G =
\frac{1}{4!} \vev{\varphi (r) \varphi (0)}_G^4 = \frac{1}{4!\,\N_1^8}
\frac{1}{r^{4(d-2)}}
\end{equation}
implies
\begin{equation}
\frac{1}{\N_4^2} = \frac{1}{4! \N_1^8} =  \frac{1}{4! 2^8 \pi^{2d}}
\Gamma \left(\frac{d-2}{2}\right)^4\,.\label{eq:appendix-Gauss-N4}
\end{equation}

\subsection{Three-point functions}

Let us denote $x_{ij}^{2}=\left|x_{i}-x_{j}\right|^{2}$, and
consider the three-point function
\begin{equation}
f_3 \left(x_{1},x_{2},x_{3}\right) \equiv
\frac{1}{\left(x_{12}^{2}\right)^{d/2-\nu_{3}}
  \left(x_{23}^{2}\right)^{d/2-\nu_{1}}
  \left(x_{13}^{2}\right)^{d/2-\nu_{2}}}
\label{eq:appendix-f-function}
\end{equation}
appearing in the three-point function
(\ref{eq:CFT-three-point-function}) of a CFT. We denote by $\tilde{f}_3$
the Fourier transform of $f_3$:
\[
f_3 \left(x_{1},x_{2},x_{3}\right)  =  \int_{p_{1}p_{2}p_{3}}
e^{-ip_{1}x_{1}-ip_{2}x_{2}-ip_{3}x_{3}} \left(2\pi\right)^{d} \delta
\left(p_{1}+p_{2}+p_{3}\right) \tilde{f}_3
\left(p_{1},p_{2},p_{3}\right)\,. 
\]
Since $f_3$ depends only on the differences $x_1 - x_3, x_1 - x_2$, we
obtain
\begin{align}
\tilde{f}_3 \left(p_{1},p_{2},-p_{1}-p_{2}\right) 
& =  \int_{x_{1},x_{2}}\,e^{ip_{1}x_{1}+ip_{2}x_{2}}
  f_3 \left(x_{1},x_{2},0\right)\nonumber   \\ 
& = \int_{x_{1},x_{2}}\,e^{i p_{1}x_{1}+i p_{2}x_{2}}
  \frac{1}{\left(x_{12}^{2}\right)^{d/2-\nu_{3}}
  \left(x_{2}^{2}\right)^{d/2-\nu_{1}}
  \left(x_{1}^{2}\right)^{d/2-\nu_{2}}}\,.
\label{eq:f-tilde-via-f_x1-x2}
\end{align}
We can express $1/\left(x_{12}^{2}\right)^{d/2-\nu_{3}}$ by using
(\ref{eq:Four-of-normalized-2pt-function}). This gives
\[
\tilde{f}_3 \left(p_{1}, p_{2}, -p_{1}-p_{2}\right) 
 =  4^{\nu_{1}+\nu_{2}+\nu_{3}}\pi^{3d/2}
  \frac{\Gamma\left(\nu_{3}\right) \Gamma \left(\nu_{1}\right)
  \Gamma\left(\nu_{2}\right)}{\Gamma\left(\frac{d}{2}-\nu_{3}\right)
  \Gamma \left(\frac{d}{2}-\nu_{1}\right)
  \Gamma\left(\frac{d}{2}-\nu_{2}\right)}
J_{\nu_1 \nu_2 \nu_3} (p_1, p_2)
\]
where
\[
J_{\nu_{1}\nu_{2}\nu_{3}}\left(p_{1},p_{2}\right)  =
\int_{q}\frac{1}{\left(q^{2}\right)^{\nu_{3}}
  \left(\left(q+p_{1}\right)^{2}\right)^{\nu_{2}}
  \left(\left(q-p_{2}\right)^{2} \right)^{\nu_{1}}}\,. 
\]
Hence,  we obtain
\begin{align*}
f_3 (x_1, x_2, x_3) 
& =  4^{\nu_{1}+\nu_{2}+\nu_{3}}\pi^{3d/2}
  \frac{\Gamma\left(\nu_{1}\right) \Gamma\left(\nu_{2}\right)
  \Gamma\left(\nu_{3}\right)}{\Gamma\left(\frac{d}{2}-\nu_{1}\right)
  \Gamma\left(\frac{d}{2}-\nu_{2}\right) \Gamma\left(\frac{d}{2}-\nu_{3}\right)}\\
 &\quad  \times\int_{p_{1}p_{2}p_{3}}\left(2\pi\right)^{d} \delta
   \left(p_1+p_2+p_3\right)
   e^{i \left( p_1 x_1 + p_2 x_2 + p_3 x_3\right)} J_{\nu_{1}\nu_{2}\nu_{3}} \left(p_{1},p_{2}\right)\,,
\end{align*}

We now consider the limit $p_{1}\gg p_{2}$.  We obtain
\begin{align*}
J_{\nu_{1}\nu_{2}\nu_{3}}\left(p_{1},p_{2}\right) 
& \overset{p_1 \gg p_2}{\longrightarrow}
  \left(p_{1}^{2}\right)^{-\nu_{2}}\int_{q}
  \frac{1}{\left(q^{2}\right)^{\nu_{3}} \left(\left(q-p_{2}\right)^{2}
  \right)^{\nu_{1}}}\\ 
 & =  \frac{1}{p_{1}^{d+\Delta_{2}-\Delta_{1}-\Delta_{3}}}
   \frac{1}{p_{2}^{d-2\Delta_{2}}} \frac{1}{\left(4\pi\right)^{d/2}}
   \frac{\Gamma \left(\nu_{1}+\nu_{3}-\frac{d}{2}\right)}{\Gamma
   \left(\nu_{1}\right) \Gamma\left(\nu_{3}\right)} \frac{\Gamma
   \left(\frac{d}{2}-\nu_{1}\right) \Gamma
   \left(\frac{d}{2}-\nu_{3}\right)}{\Gamma \left(d-\nu_{1}-\nu_{3}\right)}\,.
\end{align*}
Hence, in the same limit the Fourier transform $\tilde{f}_3$ is obtained as
\begin{align}
\tilde{f}_3 \left(p_{1},p_{2},-p_{1}-p_{2}\right) 
&\overset{p_1 \gg p_2}{\longrightarrow}  
\left(4\pi\right)^{d}4^{-\frac{1}{2} \left(\Delta_{1}+ \Delta_{2}
  +\Delta_{3}\right)} \frac{\Gamma\left(\frac{1}{2} 
  \left(d+\Delta_{2}-\Delta_{1}-\Delta_{3}\right)\right)}{\Gamma\left(\frac{1}{2}
  \left(\Delta_{1}+\Delta_{3}-\Delta_{2}\right) \right)}
  \frac{\Gamma\left( \frac{d}{2}-\Delta_{2}\right)}{\Gamma
  \left(\Delta_{2}\right)}\notag \\
&\qquad \times \frac{1}{p_{1}^{d+\Delta_{2}-\Delta_{1}-\Delta_{3}}}
  \frac{1}{p_{2}^{d-2\Delta_{2}}}\,.
\end{align}

\section{The fixed point effective average action to the first order in $\ep$
\label{App:The-fixed-point-EAA-1st-order}}

We wish to construct an ERG trajectory parametrized by $g$ along which
the Gaussian fixed point at $g=0$ is connected to the Wilson-Fisher
fixed point at $g=g_*$.  (Please see \cite{Dutta:2020vqo} for more
detailed discussions of the fixed point in $\ep$ expansions.)  The ERG
differential equation in the dimensionless framework is given by
\begin{align}
  \left( \ep g + \beta (g) \right) \partial_g \Gamma (g) [\varphi]
  &= \int_p \left( p \cdot \partial_p + \frac{d+2}{2} - \frac{\eta (g)}{2}
    \right) \varphi (p) \frac{\delta \Gamma (g) [\varphi]}{\delta
    \varphi (p)}\notag\\
  &\quad - \int_p \left(2 - p \cdot \partial_p - \eta (g) \right)
    R(p) \, \frac{1}{2} G (g)_{p, -p} [\varphi]\,,
    \label{eq:appendix-ERGdiffeq}
\end{align}
where $G(g)_{p,-q} [\varphi]$ is defined by
\begin{equation}
\int_q \left( \frac{\delta^2 \Gamma (g) [\varphi]}{\delta \varphi (p)
    \delta \varphi (-q)} + R (p) (2 \pi)^d \delta (p-q) \right) G
(g)_{q, -r} [\varphi] = (2 \pi)^d \delta (p-r)\,.
\end{equation}
We expand $\Gamma (g)$ in powers of $\varphi$ as
\begin{align}
  \Gamma (g) [\varphi]
  &= \frac{1}{2} \int_p \varphi (p) \varphi
    (-p)\, v^{(2)} (g; p) \notag\\
  &\quad + \frac{1}{4!} \int_{p_1,\cdots,p_4} \prod_{i=1}^4 \varphi
    (p_i)  (2\pi)^d \delta \left(\sum_{i=1}^4
    p_i\right)\, v^{(4)} (g; p_1,\cdots, p_4)  + \cdots\,.
\end{align}
The anomalous dimension $\frac{1}{2} \eta (g)$ of $\varphi$ in
(\ref{eq:appendix-ERGdiffeq}) is determined by the condition
\begin{equation}
  \frac{\partial}{\partial p^2} v^{(2)} (g; p)\Big|_{p^2=0} = 1\,.
  \label{appB-normalization-v2}
\end{equation}
The beta function $\beta (g)$ is determined by the boundary condition
\begin{equation}
  v^{(4)} (g; 0,0,0,0) = g\,.\label{appB-normalization-v4}
\end{equation}
The Gaussian fixed point is given by
\begin{equation}
  \Gamma (0) [\varphi] = \frac{1}{2} \int_p \varphi (p) \varphi (-p)\,
  p^2\,.
\end{equation}

In the main text we only need the first order approximation to $\Gamma
(g)$:
\begin{subequations}
\begin{align}
  v^{(2)} (g; p)
  &\simeq p^2 + g \,v_1^{(2)} (p)\,,\\
  v^{(4)} (g; p_1, p_2, p_3, p_4)
  &\simeq g \left( 1 + g \,v_2^{(4)} (p_1, p_2, p_3, p_4)\right)\,,\\
  v^{(2n \ge 6)} (g; p_1, \cdots, p_{2n})
  &\simeq \mathrm{O} (g^n)\,.
\end{align}
\end{subequations}
We expand
\begin{subequations}
  \begin{align}
    \beta (g)
    &= \beta_2 g^2 + \cdots\,,\\
    \eta (g)
    &= \eta_1 g + \cdots\,.
  \end{align}
\end{subequations}
$v_1^{(2)} (p)$ satisfies
\begin{equation}
  \left(-2 + \ep \right) v_1^{(2)} (p) = \frac{1}{2} \int_q f(q) -
  \frac{1}{2} \eta_1 p^2\,,
\end{equation}
where
\begin{equation}
  f(q) \equiv \frac{(2 - q \cdot \partial_q) R(q)}{\left(q^2 + R(q)\right)^2}\,.
\end{equation}
(\ref{appB-normalization-v2}) gives
$\eta_1 = 0$, and we obtain
\begin{equation}
  v_1^{(2)} = - \frac{1}{2-\ep} \frac{1}{2} \int_q f(q)\,,\label{eq:appendix-v12}
\end{equation}
which is a mass term independent of $p$.  $v_2^{(4)}$ satisfies
\begin{align}
&  \left( \sum_{i=1}^4 p_i \cdot \partial_{p_i} + \ep \right) v_2^{(4)}
  (p_1, p_2, p_3, p_4) \notag\\
  &= - \beta_2 - \int_q f(q) \left( h(q+p_1+p_2) + h (q+p_1+p_3) + h(q
    + p_1 + p_4) \right)\,,
\end{align}
where
\begin{equation}
  h(q) \equiv \frac{1}{q^2 + R(q)}\,.
\end{equation}
(\ref{appB-normalization-v4}) gives
\begin{equation}
  \beta_2 = - 3 \int_q f(q) h(q)\,.\label{eq:Appendix-def-beta2}
\end{equation}
We then obtain
\begin{equation}
  v_2^{(4)} (p_1,p_2,p_3,p_4) = - \lb F(p_1+p_2) + F(p_1+p_3) + F
  (p_1+p_4)\rb\,,\label{eq:appendix-v24}
\end{equation}
where $F(p)$ is defined by
\begin{equation}
  F(p) \equiv \frac{1}{2} \int_q h(q) \left( h (q+p) - h (q)
  \right)\,.
\end{equation}

The fixed point value $g_*$ is obtained from
\begin{equation}
0 =  \ep g_* + \beta (g_*) \simeq \ep g_* + \beta_2 g_*^2
\end{equation}
as
\begin{equation}
  g_* \simeq - \frac{\ep}{\beta_2} = \frac{\ep}{3 \int_q f(q) h(q)}
  \simeq \frac{(4 \pi)^2}{3} \ep\,.
\end{equation}
The value of $\beta_2$ at $\ep=0$ is calculated in (\ref{eq:Appendix-beta2}).

\section{Asymptotic behaviors of $F(p)$ and $G(p)$\label{App:asymptotic-behaviors}}

\subsection{$F(p)$}

$F(p)$ is defined by 
\begin{equation}
\left( p \cdot \partial_p + \ep \right) F(p) = \int_q f(q) \left(
  h(q+p) - h(q)\right)\,.
\end{equation}
Since $f(q)$ vanishes rapidly for $q$ beyond the cutoff scale
(order $1$), the above gives, for $p \gg 1$,
\begin{equation}
\left( p \cdot \partial_p + \ep \right) F (p) = \frac{1}{3} \beta_2 +
\mathrm{O} \left(\frac{1}{p^2}\right)\,,
\end{equation}
where $\beta_2$ is given by (\ref{eq:Appendix-def-beta2}).

This implies the asymptotic behavior
\begin{equation}
F(p) \overset{p \gg 1}{\longrightarrow} F_{\mathrm{asymp}} (p) \equiv
\frac{1}{\ep} \frac{1}{3} \beta_2  (\ep)+ C_F (\ep) p^{- \ep}\,,
\end{equation}
where we have indicated the $\ep$-dependence of $\beta_2$.
Since $F(p)$ is finite as $\ep \to 0+$, we must find
\begin{equation}
C_F (\ep) \overset{\ep \to 0+}{\longrightarrow} - \frac{1}{\ep}
\frac{1}{3} \beta_2 (\ep=0) + \mathrm{O} (\ep^0)\,.
\end{equation}

To compute $C_F (\ep)$ exactly, we recall the high momentum propagator
has the same short-distance singularity in coordinate space as the Gaussian two-point
function:
\begin{equation}
\tilde{h} (r) \equiv \int_p e^{i p r} h(p) \overset{r \to
  0}{\longrightarrow} \int_p \frac{e^{i pr}}{p^2} = \vev{\varphi (r)
  \varphi (0)}_G = \frac{1}{4
  \pi^{\frac{d}{2}}} \Gamma \left(\frac{d-2}{2}\right)
\frac{1}{r^{d-2}}\,.
\end{equation}
Since
\begin{equation}
F(p) = \frac{1}{2} \int_q h(q)  \left( h(q+p) - h(q)\right)\,,
\end{equation}
we obtain
\begin{equation}
\tilde{F} (r) \equiv \int_p F(p) e^{i p r} = \frac{1}{2} \tilde{h}
(r)^2 \overset{r \to 0}{\longrightarrow} \frac{1}{2} \vev{\varphi (r)
  \varphi (0)}_G^2\,.
\end{equation}
Thus, the $p$-dependent part of the asymptotic behavior of $F(p)$ is
the same as the Fourier transform of the squared Gaussian propagator.
Hence, using (\ref{eq:Four-of-normalized-2pt-function}), we obtain
\begin{equation}
C_F (\ep) = \frac{1}{2} \frac{1}{(4 \pi)^{\frac{d}{2}}} \frac{\Gamma
  \left(\frac{d-2}{2}\right)^2 \Gamma \left(2 -
    \frac{d}{2}\right)}{\Gamma (d-2)}
\overset{\ep \to 0}{\longrightarrow} \frac{1}{\ep} \frac{1}{(4 \pi)^2}\,.
\end{equation}
This implies
\begin{equation}
\frac{1}{3} \beta_2 (0) = - \frac{1}{(4 \pi)^2}\,.\label{eq:Appendix-beta2}
\end{equation}

In the main text we need the expansion of the asymptotic behavior to
order $\ep$:
\begin{align}
F_{\mathrm{asymp}} (p)
&= \frac{1}{2 (4 \pi)^2} \left[ - \ln \frac{p^2}{4} + \ln \pi - \gamma
  + 2 + \frac{2}{3} (4 \pi)^2 \beta_2' (0) \right.\\
&\quad\left. + \ep \lb \frac{1}{4} \left(\ln
  \frac{p^2}{4\pi}\right)^2 + \left(\frac{\gamma}{2} - 1 - \frac{1}{2}
  \ln \pi \right) \ln \frac{p^2}{4} + \mathrm{const} \rb
  \right]\,.\label{eq:Appendix-Fasymp}
\end{align}
$\beta'_2 (\ep = 0)$ depends on the choice of the cutoff function
$R(p)$.

\subsection{$G(p)$}

Similarly, we can obtain the asymptotic behavior of $G(p)$, which is
defined by
\begin{align}
\left( p \cdot \partial_p - 2 + 2 \ep \right) G(p) 
&= \int_q f(q) F(q+p) + v^{(2)}_1 \frac{2}{3} \beta_2 + \eta_2\,
  p^2\notag\\
&\overset{p \to \infty}{\longrightarrow} \eta_2\, p^2 + \mathrm{O} (p^0)\,,
\end{align}
where
\begin{subequations}
\begin{align}
v^{(2)}_1 &= - \frac{1}{2-\ep} \frac{1}{2} \int_q f(q)\,,\\
\eta_2 &= - \frac{d}{dp^2} \int_q f(q) F(q+p)\Big|_{p=0}\,.
\end{align}
\end{subequations}
The equation gives the asymptotic behavior
\begin{equation}
  G (p) \overset{p \gg 1}{\longrightarrow} G_{\mathrm{asymp}} (p)
  \equiv  \frac{1}{2 \ep}  \eta_2 (\ep)  p^2 + C_G (\ep) p^{2 - 2 \ep}\,.
\end{equation}
Since $G(p)$ is finite as $\ep \to 0+$, this implies
\begin{equation}
C_G (\ep) \overset{\ep \to 0+}{\longrightarrow} -
\frac{\eta_2 (\ep=0)}{2 \ep} \,.
\end{equation}

We can calculate $C_G (\ep)$ exactly from
\begin{equation}
\tilde{G} (r) \equiv \int_p G(p) e^{i p r} = \frac{1}{3!} \tilde{h}
(r)^3 \overset{r \to 0}{\longrightarrow} \frac{1}{3!} \vev{\varphi (r)
  \varphi (0)}_G^3\,.
\end{equation}
Using  (\ref{eq:Four-of-normalized-2pt-function}), we obtain
\begin{equation}
C_G (\ep) = \frac{1}{3!} \frac{1}{(4 \pi)^d} \frac{\Gamma
  \left(\frac{d-2}{3}\right)^3 \Gamma (3 - d)}{\Gamma
  \left(\frac{3}{2} (d-2)\right)} \overset{\ep \to 0}{\longrightarrow}
- \frac{1}{\ep} \frac{1}{12 (4 \pi)^4}\,.
\end{equation}
Hence,
\begin{equation}
\eta_2 (0) = \frac{1}{6} \frac{1}{(4 \pi)^4}\,.
\end{equation}

In the main text we need $G_{\mathrm{asymp}} (p)$ to order $\ep^0$:
\begin{equation}
G_{\mathrm{asymp}} (p) = p^2 \left(  \frac{1}{2 \cdot 3!} \frac{1}{(4 \pi)^4} \ln p^2
+ \textrm{const} \right)\,.\label{eq:Appendix-Gasymp}
\end{equation}

%%%%%%%%%%%%%%%%%%%%%%%%%%%%%%%%%%%%%%%%%%%%%%%%%%%%%%%%%%

\end{document}